\pdfoutput=1
\def\preprintbuild{}

\ifdefined\preprintbuild
  \documentclass[review=false]{jfp-epi}
\else
  \documentclass[review]{jfp-epi}
\fi

\usepackage{booktabs}
\usepackage{subcaption}
\usepackage{balance}
\usepackage{pifont}
\usepackage{wrapfig}
\usepackage{hyperref}
\usepackage{fancyvrb}
\usepackage{xspace}
\usepackage{algorithmicx}
\usepackage{algpseudocode}
\usepackage{algorithm}
\usepackage{amsfonts}
\usepackage{amsmath}
\usepackage{amsthm}
\usepackage{semantic}
\usepackage{listings}
\usepackage{array}
\usepackage{mathpartir}
\usepackage{stmaryrd}
\usepackage{graphicx}
\usepackage{syntax, etoolbox}
\usepackage{tikz}
\usepackage{proof}
\usepackage{color}
\usepackage{mathtools}
\usepackage{adjustbox}
\usepackage{multirow}
\usepackage{bbding}
\usepackage{soul}
\usepackage{colortbl}

\graphicspath{{fig/}}

\theoremstyle{definition}

\lstdefinestyle{ruler}{
  basicstyle=\footnotesize\ttfamily,
}
\lstdefinestyle{rewrite}{
  basicstyle=\scriptsize\sffamily,
  gobble=4,
}

\ifdefined\preprintbuild\else\input{drafting}\fi

\definecolor{bp}{rgb}{0.2, 0.2, 0.6}

\newcommand{\sexp}{s-expression\xspace}
\newcommand{\sexps}{s-expressions\xspace}
\newcommand{\enumo}{\textsc{Enumo}\xspace}

\newcommand{\Union}{\textsf{Union}\xspace}
\newcommand{\Filter}{\textsf{Filter}\xspace}
\newcommand{\Plug}{\textsf{Plug}\xspace}

\newcommand{\eggl}{$\mathsf{egraph}$\xspace}
\newcommand{\ruleset}{$\mathsf{ruleset}$\xspace}
\newcommand{\rl}{$\mathsf{rule}$\xspace}

\newcommand{\toeggl}{$\mathsf{to\_egraph}$\xspace}

\newcommand{\ruler}{Ruler\xspace}
\newcommand{\rust}{Rust\xspace}

\newcommand{\herbie}{Herbie\xspace}
\newcommand{\sz}{Szalinski\xspace}

\newcommand{\Ruler}{Ruler\xspace}
\newcommand{\egg}{\texorpdfstring{\MakeLowercase{\texttt{egg}}}{\texttt{egg}}\xspace}
\newcommand{\egraph}{\mbox{e-graph}\xspace}
\newcommand{\egraphs}{\mbox{e-graphs}\xspace}
\newcommand{\Egraph}{\mbox{E-graph}\xspace}
\newcommand{\Egraphs}{\mbox{E-graphs}\xspace}

\newcommand{\eclass}{\mbox{e-class}\xspace}
\newcommand{\eclasses}{\mbox{e-classes}\xspace}
\newcommand{\ematching}{\mbox{e-matching}\xspace}
\newcommand{\Ematching}{\mbox{E-matching}\xspace}

\newcommand{\enode}{\mbox{e-node}\xspace}
\newcommand{\enodes}{\mbox{e-nodes}\xspace}

\newcommand{\cvec}{cvec\xspace}

\newcommand{\Eqsat}{Equality saturation\xspace}
\newcommand{\eqsat}{equality saturation\xspace}
\newcommand{\ff}{fast-forwarding\xspace}

\newcommand{\lhs}{\T{LHS}\xspace}
\newcommand{\lhsandrhs}{\T{LHS-RHS}\xspace}

\newcommand{\denote}{$\llbracket \cdot \rrbracket$\xspace}

\newcommand{\C}[1]{\text{\footnotesize{\texttt{#1}}}\xspace}

\newcommand{\W}{\ensuremath{\mathcal{W}}}
\newcommand{\N}{$\mathbb{N}$}

\newcommand{\T}[1]{\texttt{#1}}

\newcommand{\foottt}[1]{{\footnotesize \texttt{#1}}}
\newcommand{\cis}{\mathrm{cis}}

\newcommand\rewritesto{\ensuremath{\rightsquigarrow}}
\newcommand\rewritesboth{\ensuremath{\leftrightsquigarrow}}

\begin{document}

\title{Equality saturation theory exploration à la carte}

\author{Anjali Pal}
\orcid{0009-0006-0692-0707}
\affiliation{%
  \institution{University of Washington}
  \city{Seattle}
  \country{USA}
  \authoremail{anjalip@cs.washington.edu}
}

\author{Brett Saiki}
\orcid{0009-0002-3482-5767}
\affiliation{%
  \institution{University of Washington}
  \city{Seattle}
  \country{USA}
  \authoremail{bsaiki@cs.washington.edu}
}

\author{Ryan Tjoa}
\orcid{0009-0003-0731-5398}
\affiliation{%
  \institution{University of Washington}
  \city{Seattle}
  \country{USA}
  \authoremail{rtjoa@cs.washington.edu}
}

\author{Cynthia Richey}
\orcid{0009-0008-4456-9406}
\affiliation{%
  \institution{University of Pennsylvania}
  \city{Philadelphia}
  \country{USA}
  \authoremail{lapwing@seas.upenn.edu}
}

\author{Amy Zhu}
\orcid{0000-0001-5766-7090}
\affiliation{%
  \institution{University of Washington}
  \city{Seattle}
  \country{USA}
  \authoremail{amyzhu@cs.washington.edu}
}

\author{Oliver Flatt}
\orcid{0000-0002-0656-235X}
\affiliation{%
  \institution{University of Washington}
  \city{Seattle}
  \country{USA}
  \authoremail{oflatt@cs.washington.edu}
}

\author{Max Willsey}
\orcid{0000-0001-8066-4218}
\affiliation{%
  \institution{University of California, Berkeley}
  \city{Berkeley}
  \country{USA}
  \authoremail{mwillsey@berkeley.edu}
}

\author{Zachary Tatlock}
\orcid{0000-0002-4731-0124}
\affiliation{%
  \institution{University of Washington}
  \city{Seattle}
  \country{USA}
  \authoremail{ztatlock@cs.washington.edu}
}

\author{Chandrakana Nandi}
\orcid{0000-0001-8633-8413}
\affiliation{%
  \institution{Certora Inc.}
  \city{Seattle}
  \country{USA}
  \authoremail{chandra@certora.com}
}

\renewcommand{\shortauthors}{%
  A. Pal, B. Saiki, R. Tjoa, C. Richey, A. Zhu, O. Flatt,
  M. Willsey, Z. Tatlock, C. Nandi}

\begin{abstract}
  Rewrite rules are critical in \eqsat,
  an increasingly popular technique in
  optimizing compilers, synthesizers, and verifiers.
Unfortunately, developing high-quality rulesets
  is difficult and error-prone.
Recent work to automatically infer rewrite rules
  does not scale to large terms or grammars.
Users struggle to guide inference and
  incrementally construct rulesets because
  existing rule inference tools
  are monolithic and opaque.
As a result,
  most \eqsat users still
  manually develop and maintain rulesets.

This paper proposes \enumo,
  a new domain-specific language for
  \textit{programmable theory exploration}.
\enumo provides a small set of core operators that
  enable users to strategically guide rule inference
  and incrementally build rulesets.
Short \enumo programs easily replicate
  results from state-of-the-art tools like \Ruler, but
  \enumo programs can also scale to infer
  deeper rules from larger grammars
  than prior approaches.
\enumo's composable operators even facilitate
  developing new strategies for ruleset inference.
We introduce a new \textit{\ff} strategy
  which does not require evaluating terms in the target language,
  and thus supports domains that
  were out of scope for prior work.
\enumo is also easy to extend:
  two new operators suffice to incorporate
  large language models into rule inference,
  where they complement guided search.

We evaluate \enumo and \ff across a variety of domains.
Compared to state-of-the-art techniques,
  \enumo can synthesize better rulesets
  over a diverse set of domains,
  in some cases matching the effects of
  manually developed rulesets in
  systems driven by \eqsat.

\end{abstract}

\maketitle

\AtBeginEnvironment{grammar}{\small}
\section{Introduction}
\label{sec:intro}

Equational theories in the form of rewrites ($\ell \rewritesto r$)
 have long been used in term rewriting systems.
Equality saturation engines in particular,
 which have seen a recent resurgence,
 leverage these theories to power systems in a wide
  variety of domains like program synthesis~\citep{szalinski, herbie, spores, babble, resys},
  formal verification~\citep{coq, z3, sat19, rest-darulova, sam1, sam2}, and
  optimizing compilers~\citep{eqsat, tensat, denali, wetune, deepegg, qgym, sketch-eq}.
A principal challenge in building these systems
 is writing the rewrites themselves:
 too few rewrites can lead to missed optimizations;
 too many can make implementation and maintenance difficult.
 Furthermore, even one incorrect rewrite can compromise the soundness
 of the entire system.

\textit{Theory explorers} automatically generate
 equational theories~\citep{sat19, ruler, cav21, hipster, quickspec, hipspec, isacosy}.
These tools generally follow a three-stage approach:
\begin{enumerate}
  \item
    Enumerate terms from a given grammar,
    typically in a bottom-up,
    exhaustive manner~\citep{ruler, cvc4}.
  \item
    Generate candidate rewrite rules from the enumerated terms.
    Naively, any pair of enumerated terms may be a candidate rewrite rule.
    Prior work has used techniques like finger-printing, fuzzing,
    and symbolic execution to identify ``likely sound'' candidates~\citep{ruler,sat19,cav21, bansal}.
  \item
    Using the candidates,
     select a set of rewrite rules that are both sound and useful.
    Typically,
    this is done via a process that verifies the candidates
    and removes redundant ones.
   \citet{ruler} call this process ``minimization,''
    under the assumption
    that a smaller set of rules is more likely to be effective.
\end{enumerate}

Despite recent innovations,
  theory explorers are still not widely used.
We posit that their monolithic
 implementations make them
 too inflexible.
They are designed for idealized ``one-shot'' use cases:
 the user provides a grammar, interpreter, and verifier,
 presses a button, and out comes a ruleset (set of rewrites)
 ready for use in a rewriting or equality saturation-based system.
In reality,
 tools based on equational theories
 are not developed or maintained in this manner.
Instead,
 engineers and domain experts
 build, maintain, measure, debug, and compare rulesets
 both \emph{iteratively} over time
 and \emph{incrementally} as new features and requirements are added.
In addition,
 automated theory explorers are often intended
 to replace or augment existing (handwritten) rulesets,
 but their rigid, one-shot approach
 leaves developers with little recourse
 when the output is not 100\% satisfactory.

\begin{figure}
  \centering\includegraphics[width=0.75\linewidth]{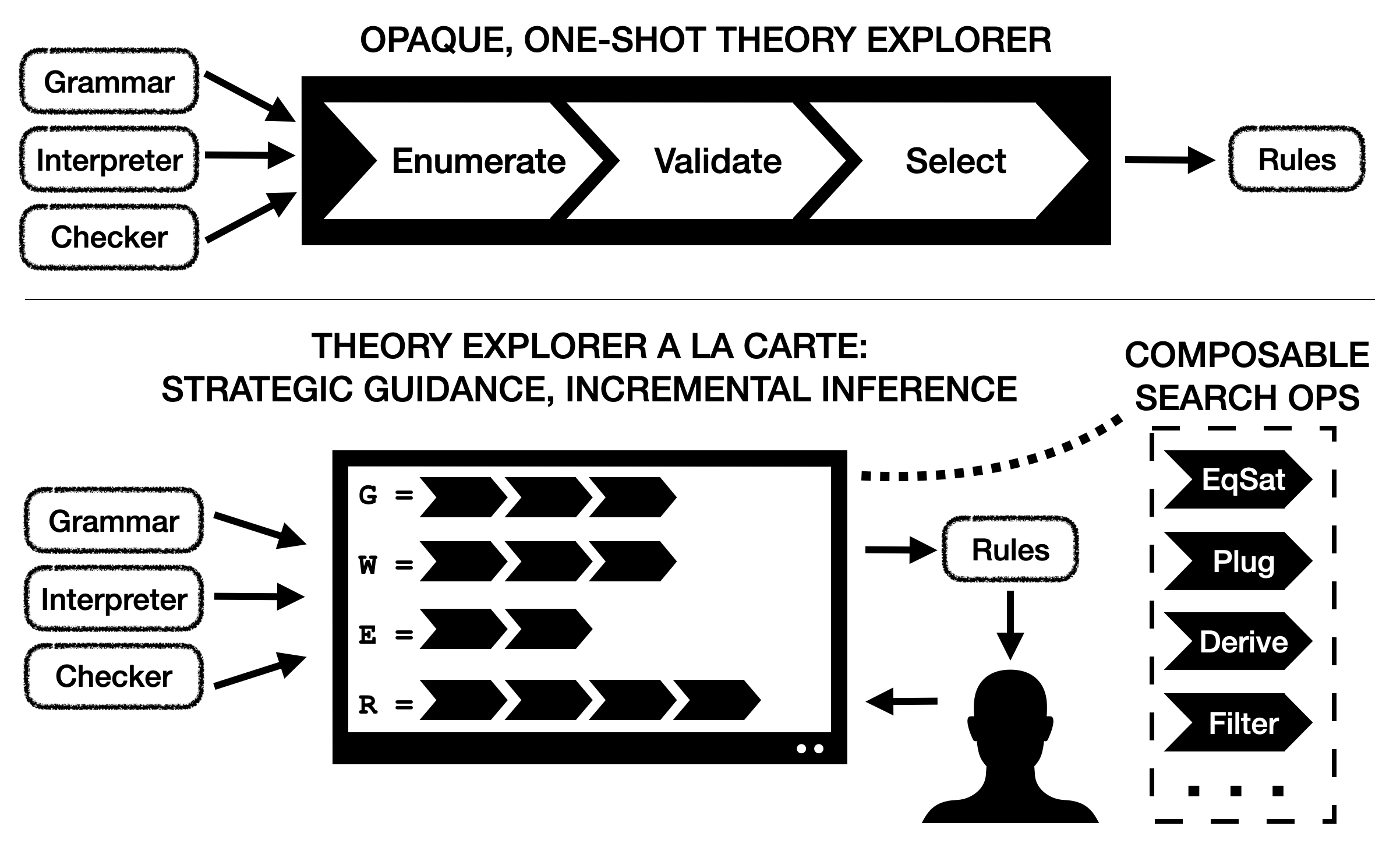}
  \caption{
    \textit{(Top)}
    Typical theory explorer workflow:
      the user provides a grammar, interpreter, and rule validator
      and gets a ruleset.
    Such theory explorers are rigid and opaque;
      there is no mechanism for users to intervene
      or apply domain expertise to guide inference.
    \textit{(Bottom)}
    In contrast, the \enumo DSL
      enables the user to guide theory exploration.
    Users provide the same three inputs
      as well as a short \enumo program.
    \enumo's modular and composable operators
      make it easy to implement existing inference strategies,
      add domain-specific tweaks, or even implement new strategies.
      \vspace{-0.75em}
  }\label{fig:workflow}
\end{figure}

Existing theory explorers do not scale to the
  needs of real systems.
For example,
  $x + y \rewritesto \frac{x^2 - y^2}{x-y}$
  is useful for factoring in numerical applications.
However,
  discovering it via exhaustive enumeration,
  as used in recent approaches~\citep{sat19, ruler},
  is infeasible within even a moderately sized grammar.
Instead,
  users should be able to help guide
  the theory explorer to discover such rules.

We present a new paradigm:
  theory exploration \emph{\`a la carte},
  which breaks theory explorers down
  into a set of modular operators.
Users can programmatically compose these operators
 to easily build a theory explorer suited to their needs.
We developed \enumo,
  an embedded domain-specific language (DSL)
 in which term enumeration strategies and
 rulesets are first-class values.
Simple \enumo programs can
  generate useful rulesets that
  prior work~\citep{ruler} cannot.
\enumo's abstractions also
  inspired ``fast-forwarding,''
  a new theory exploration algorithm
  that supports domains where
  equality is undecidable (e.g., real arithmetic).

We demonstrate that \enumo
  programs can synthesize
  better rulesets compared
  to state-of-the-art tools,
  while also scaling to
  much larger grammars.
In a case study inspired by Halide's large grammar~\citep{halide},
  an \enumo program synthesized a
  ruleset that derives
  90\% of Halide's handwritten rules.
Compared to prior work in theory exploration,
  fast-forwarding enabled Herbie~\citep{herbie}
  to achieve 35\% higher accuracy when
  improving floating-point rounding error (\autoref{subsec:eval-ff})
and allowed us to find alternate implementations of trigonometric
  functions in Megalibm, another floating-point synthesis tool~\citep{megalibm}.
Finally, in the domain of constructive solid geometry,
  \enumo's synthesized rules for CAD identities allow \sz~\citep{szalinski}
  to shrink benchmarks by 87\% on average,
  closely matching the 90\% reduction achieved by
  expert-written rules.

In summary, this paper makes the following contributions:
\begin{itemize}
  \item A DSL, \enumo, that offers operators for
    generating custom workloads,
    composing theory exploration strategies,
    and manipulating rulesets (\autoref{sec:slide}).
  \item A new algorithm for ``fast-forwarding'' rules to
    infer rulesets in domains where providing
    an interpreter is infeasible (\autoref{sec:lift}).
  \item An extensive evaluation showing that,
      compared to a state-of-the-art theory explorer,
      custom workloads and ruleset composition lead
      to better rulesets (\autoref{sec:eval}).
  \item A set of end-to-end case studies demonstrating that
    \enumo's synthesized rulesets are comparable to handwritten rulesets
    across a variety of domains (\autoref{sec:eval}).
  \item A case study extending \enumo with two operators that
    incorporate large language models (LLMs) into rule and workload
    synthesis, showing that the new operators compose with \enumo's
    existing pipeline and that LLM-based and guided-search rules
    are complementary
    (\autoref{sec:case-study-llm}).
  \item A detailed account of the developer experience of
    incrementally building rulesets with \enumo (\autoref{sec:devexp}).
\end{itemize}

This paper extends our earlier conference
version~\citep{enumo-oopsla23} with an expanded treatment of
derivability and proving power (\autoref{sec:slide},
\autoref{sec:eval}), a complete presentation of fast-forwarding
together with a trigonometric case study (\autoref{sec:lift},
\autoref{sec:devexp}), and a new case study extending \enumo with
operators that incorporate LLMs (\autoref{sec:case-study-llm}).

\section{Background on \eqsat}
\label{sec:back}

\begin{figure}[t]
\footnotesize
  \begin{minipage}{0.55\linewidth}
    \begin{lstlisting}[xleftmargin=0pt]
    def equality_saturation($t$, $R$):
      egraph = empty_egraph()
      $c_{\sf root}$ = egraph.add($t$)                       $\label{ln:eqsat-init}$
      saturated = False
      until saturated or timeout():
        saturated = True
        for $\ell \rewritesto r$ in $R$:
          for ($\sigma$, $c_\ell$) in egraph.search($\ell$): $\label{ln:eqsat-ematch}$
            $c_r$ = egraph.add($\sigma(r)$)                  $\label{ln:eqsat-add}$
            if not egraph.same_eclass($c_\ell$, $c_r$):
              egraph.union($c_\ell$, $c_r$)                  $\label{ln:eqsat-merge}$
              saturated = False
      return egraph.extract_best($c_{\sf root}$)             $\label{ln:eqsat-extract}$
  \end{lstlisting}
  \end{minipage}
  \begin{minipage}{0.44\linewidth}
    \begin{center}
    \includegraphics[scale=0.5]{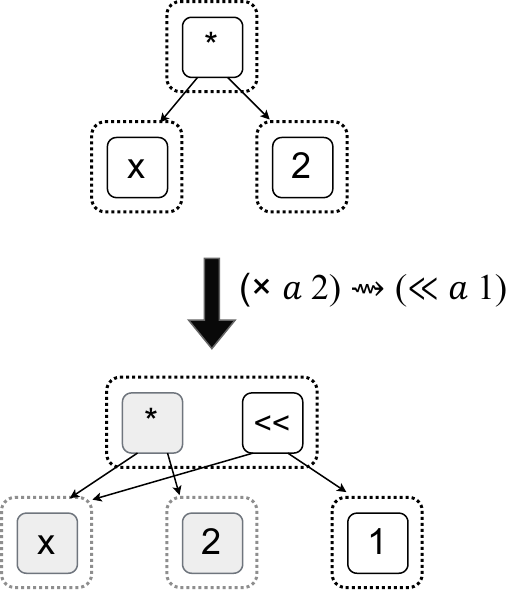}
    \end{center}
  \end{minipage}
  \caption{
    (Left) The \eqsat algorithm~\citep{eqsat, egg, ruler}. Initially, a
    new \egraph is created that represents the input term $t$.
    Sound rewrite rules
    are applied until saturation or some resource bound (like iteration limit
    or timeout) is reached.
    A cost function is used to extract the ``best'' program from the
    \egraph.
    (Right) Examples of two \egraphs before and after applying
    the rewrite $(\times ~ a ~ 2) \rewritesto (\ll ~ a ~ 1)$.
    The dotted boxes represent \eclasses and the solid boxes represent \enodes.
    The \egraph on top represents the
    AST (abstract syntax tree) of the input term,
    $(\times ~ x ~ 2)$, and the \egraph below
    shows that the \eclasses that represent the terms $(\times ~ x ~ 2)$ and
    $(\ll ~ x ~ 1)$ are merged after the rule is applied.
    }
  \label{fig:eqsat-alg}
\end{figure}

This paper investigates strategic theory exploration powered by
 the \eqsat technique and the \egraph data structure.
Here we provide a brief background on both topics.

\subsection{E-graphs}
\label{sec:bg-egraphs}

An \egraph~\citep{nelson, kozen} is a data structure
 to efficiently represent an equivalence relation over terms.
An \egraph is a set of \emph{\eclasses},
 and each \eclass is a set of equivalent \emph{\enodes}.
An \enode $f(c_1, c_2, ...)$ is a function $f$ with children \eclasses $c_i$.

An \egraph is said to \emph{represent} a term $t$
 if any of its \eclasses represents $t$;
 an \eclass represents $t$
 if any \enode in the \eclass represents it.
An \enode $f(c_1,\ldots, c_n)$
 represents a term $f(t_1,\ldots, t_n)$
 if each $c_i$ represents $t_i$.
Two terms represented by the same \eclass
 are considered equivalent.
We additionally define some operators over
  \egraphs that are useful later in the paper.
\begin{itemize}
  \item \T{add($t$)} 
  adds term $t$ to the \egraph,
  and returns the \eclass representing $t$.
\item \T{lookup($t$)} returns
  the \eclass that represents a term $t$ if such an \eclass exists.
\item \T{merge($c_1, c_2$)} takes two \eclass ids
 and combines them into a single \eclass.
\end{itemize}
\citet{egg} give the semantics of these operators in more detail.
\Egraphs were developed for automated theorem proving
 and are today used in SMT solvers~\citep{cvc4, z3}.
More recently, \egraphs have been used
 for a program optimization technique called \eqsat~\citep{eqsat, egg}.

\subsection{Equality saturation}

Consider a term $t$ and a set of rewrite rules $R$.
A rewrite rule, $\ell \rewritesto r$, is
  a pair of patterns.
A traditional term rewriting system
 applies a rewrite
 by finding substitutions $\sigma$ such that $\sigma(\ell)$ is a subterm of $t$,
 and then replacing the subterm with $\sigma(r)$.
In this paradigm, the final output term can vary greatly depending
 on the order in which rewrites are applied.

\Eqsat~\citep{eqsat,egg} is an alternative to conventional
 term rewriting
 that uses an \egraph to \emph{non-destructively} apply rewrites.
In \eqsat, matched
  subterms are not replaced; instead they are merged into the same \eclass.

The core algorithm
  is shown in \autoref{fig:eqsat-alg}, alongside an example of an \egraph
  before and after a rewrite rule is applied.
The \eqsat algorithm takes as input a term $t$, a set of rewrite rules $R$,
 and some resource limits (e.g., timeout, iteration limit, \egraph size in terms
 of number of \enodes),
 and outputs a term $t'$ that is equivalent to $t$.
First, a new \egraph representing $t$ (line~\ref{ln:eqsat-init}) is created.
\Eqsat then applies each rewrite rule in $R$ to the \egraph.
Rule application happens in three stages:

\begin{enumerate}
  \item Line \ref{ln:eqsat-ematch}: an algorithm called
    \ematching~\citep{simplify, cade} finds all terms in the \egraph that
    match the pattern $\ell$.
    \Ematching returns a list of tuples
    $(\sigma, ec_\ell)$ where $\sigma$ is the substitution and $ec_\ell$ is an
    \eclass that represents the term $\sigma(\ell)$.
  \item Line \ref{ln:eqsat-add}: for each tuple
   $(\sigma, ec_\ell)$, \eqsat applies $\sigma$ to
   $r$ to get a term $\sigma(r)$.
  If $\sigma(r)$ is not already represented by the
    \egraph, it is added in a new
    \eclass.
  \item Line \ref{ln:eqsat-merge}:
   the \eclass representing $\sigma(\ell)$ and the \eclass
   representing $\sigma(r)$ are merged.
\end{enumerate}

Ideally, rewrites are applied until the \egraph
  saturates---i.e., no new \enodes are added to the \egraph.
In practice,
  saturation is rare,
  and resource bounds like iteration or
  \enode limits are necessary to control the size of the \egraph.
Following rule application, a cost function is used to
  extract the ``best'' expression
  equivalent to $t$ from the \egraph.

Many tools have used \eqsat for driving program transformations, program
  synthesis, and equivalence checking~\citep{eqsat, herbie, szalinski, spores, yogo, alexa}.
Recently, \eqsat has been used to find rulesets
  for other \eqsat-based systems~\citep{cav21, ruler}.

\subsection{Equational theory inference}

Theory exploration tools
  infer a set of axioms
  for a given domain.
For the purposes of this paper,
 we will only focus on equational axioms,
 which most of these tools emit.

As described in \autoref{sec:intro}, equational theory inference
  typically follows a three-step process: (1) term enumeration, (2) candidate
  generation, (3) rule filtering.
In existing theory exploration tools,
 these steps are embedded in the core synthesis algorithm,
 making it difficult or impossible
 for users to guide or customize
 the tools according to their use case.
This paper presents the \enumo DSL,
  which makes theory exploration modular by
  offering a small set of
  composable operators.

\section{\enumo by example}
\label{sec:overview}
\enumo is a DSL that provides
 the operators needed to build a
 theory explorer for \eqsat,
 driven by \eqsat, \textit{\`a la carte}.
Users define their own term enumeration strategies and
customize how the resulting rules are processed
and combined.

To introduce the basics of using \enumo,
 we will walk through a simple example
 learning rules over the domain of rational arithmetic.
 \autoref{sec:overview-basic}
 recreates prior work on theory exploration \citep{ruler}
 in a few lines of \enumo code.
\autoref{sec:guided-enumeration}
 goes beyond the capabilities of existing tools
 by employing \enumo's workload construction operators
 to guide term enumeration.
\autoref{subsec:ffoverview}
 demonstrates \enumo's ruleset manipulation primitives,
 including a new way to learn rules
 \emph{without} an interpreter.

\subsection{\enumo basics: Learning rules for rational arithmetic}
\label{sec:overview-basic}
\label{subsec:basic}
Consider the task of
  learning rules for the domain of
 rational arithmetic
 (a grammar with operators \texttt{+, -, $\times$, /, abs, $\sim$})
 for which we assume the user can provide a
 concrete evaluator (a standard recursive
 interpreter).
As with any theory exploration tool,
 the first step is to
 enumerate terms from the domain.
This is typically accomplished by giving a grammar
 to the theory explorer, which then exhaustively enumerates
 terms up to some depth.
In \enumo, however, it is up to the user
 to construct a \emph{workload} that enumerates terms
 by composing various workload primitives and combinators.
The simplest workload is a set of \sexps:
\begin{lstlisting}[name=overview]
  leaves = { a b c -1 0 1 }
  grammar = {
    EXPR                $\label{ln:bare-expr}$
    (- EXPR)
    (abs EXPR)
    (+ EXPR EXPR)
    (- EXPR EXPR)
    (* EXPR EXPR)
    (/ EXPR EXPR)
  }
\end{lstlisting}

Above, the \C{leaves} workload is a set of six atomic \sexps
 that represent the atoms of the domain:
 symbols \C{a}, \C{b}, \C{c}, and constants \C{-1}, \C{0}, and \C{1}.
The \C{grammar} workload is a set of
  \sexps that represent the grammar of the domain.
The \C{EXPR} symbol is intended as a placeholder
 but has no special meaning to \enumo.
By convention, we fully capitalize these placeholder
  symbols to differentiate them from symbols
  and operators in the domain.

The \C{plug} operator allows the user to compose two workloads
 $\W_1$ and $\W_2$ by replacing occurrences of a given symbol $x$
 in $\W_1$ with \sexps from $\W_2$.
Using \C{plug},
 the user can construct a workload that exhaustively enumerates terms
 up to a particular depth:

\begin{lstlisting}[name=overview]
  rationals_depth1 = grammar.plug("EXPR", leaves)           $\label{ln:plug1}$
  rationals_depth2 = grammar.plug("EXPR", rationals_depth1) $\label{ln:plug2}$
\end{lstlisting}

Note that \C{$\W_1$.plug("x", $\W_2$)} yields
 all possible combinations of replacing symbol \C{"x"} in $\W_1$ with an \sexp from $\W_2$
 (see \autoref{sec:slide} for detailed semantics).
Because the \C{grammar} on line \ref{ln:bare-expr} includes \C{EXPR},
\C{grammar.plug("EXPR", W)} will include all \sexps in \C{W}.
Lines \ref{ln:plug1} and \ref{ln:plug2}
 will enumerate all terms up to depth 1 and 2, respectively.

With a workload in hand
 that represents enumerated terms from the domain,
 we can proceed with writing an \enumo program to learn rules.
Initially,
 we will learn rules following
 the conventional approach
 of \citet{ruler} and \citet{sat19};
 in later subsections, we will show a more advanced \enumo program.
First, we learn rules of depth 1 (D1):

\begin{lstlisting}[name=overview]
  candidates1 = rationals_depth1
                 .to_egraph()                       $\label{ln:to-egraph}$
                 .find_candidates()                     $\label{ln:cvec-match}$
  (valid1, _) = candidates1.partition(|c| c.is_valid()) $\label{ln:validate}$
  rules1 = valid1.minimize([]) $\label{ln:minimize}$
\end{lstlisting}

Line $\ref{ln:to-egraph}$ converts the D1 workload to an \egraph
 by evaluating the workload (according to the semantics in \autoref{sec:slide})
 and adding the resulting \sexps to the \egraph.
Unlike the untyped \sexps in the workload,
 the \egraph is typed according to the domain $L$
 (in this case, rational arithmetic).\footnote{
  An exception occurs if \sexps cannot be parsed into the syntax for domain $L$.
 }
Once the \egraph is constructed,
 we use \C{find\_candidates}
 to generate rule candidates (line $\ref{ln:cvec-match}$).
This method uses the \emph{characteristic vector} (\cvec) matching
 approach from \citet{ruler},
 evaluating the terms in the \egraph
 with the user-provided interpreter on a sampling of constants.
Terms that evaluate identically on all inputs
 are likely to be \textit{equivalent}, and are thus candidates for rules.
Candidates are not necessarily sound,
 so we must validate them on line $\ref{ln:validate}$
 using a user-provided verifier
 (over the rationals, we use Z3~\citep{z3}).

Finally, on line $\ref{ln:minimize}$,
 we minimize the candidates
 to eliminate redundant rules.
The \C{minimize} operator in \enumo
 is parameterized over a scheduling \C{Strategy}.
With a \C{Strategy}, a user can control
 how much redundancy is permissible in the ruleset,
 measured via ``derivability'', as defined
 in \autoref{subsec:derivability}.
Minimize works as follows:
rules from \C{valid1} are added one by one
to a new, initially empty ruleset \C{rules1}. A given rule $r \in$ \C{valid1}
is added to \C{rules1} if the rules currently in \C{rules1} \textit{cannot}
derive $r$ within a given number of \eqsat iterations.

At this point,
 this small \enumo program has emulated the
 behavior of \citet{ruler}'s \ruler tool
 to learn D1 rules.
We can now learn rules of depth 2 (D2)
 by repeating the process:
\pagebreak

\begin{lstlisting}[name=overview]
  candidates2 = rationals_depth2 # from line (*@\ref{ln:plug2}@*)
                 .to_egraph()
                 .compress(rules1) $\label{ln:compress}$
                 .find_candidates()
  rules2 = candidates2.minimize(rules1)
  all_rules = rules2.union(rules1)
\end{lstlisting}

Again following \citet{ruler},
 learning the D2 rules benefits from the D1 rules
 in two ways.
First,
 on line $\ref{ln:compress}$,
 we
 compress the D2 \egraph
 prior to finding candidates.
This not only shrinks the \egraph
 and makes finding candidates more efficient,
 but also prevents learning candidates that are implied by the D1 rules.
Similarly,
 we pass the D1 rules to the \C{minimize} operator
 to minimize candidates not only with respect to each other,
 but also with respect to the D1 rules.
 Finally,
 we compose \C{rules2} and \C{rules1} into \C{all_rules}.

\subsection{Guided enumeration to find ``deeper'' rules}
\label{sec:guided-enumeration}

Notice that we have just implemented the entire \ruler tool in about twenty
  lines of \enumo.
\enumo workload operators can easily express exhaustive term enumeration, and
  the candidate generation and selection techniques used in \ruler are also
  supported in \enumo.
However, while \ruler can only perform a few iterations before getting stuck
  due to the exponential growth of enumerated terms, \enumo programs can
  express subsets of the entire term space, making it easy to scale beyond what
  is possible with exhaustive enumeration.

In the previous example (\autoref{sec:overview-basic}), we learned rules over
the \C{rational} domain, which is complicated by rewrites involving division (/)
that are only conditionally true---for example, the rule \C{(/ a a) \rewritesto\,1}
holds only when \C{a} is nonzero. To address this problem, we implemented
a version of the \C{rational} domain in \enumo that supports conditional statements.

Suppose we want to find the rule
  \C{(/ a a) \rewritesto\, (if a 1 (/ a a))},
  which is a version of the rule
  \C{(/ a a) \rewritesto\, 1} that inserts
  a condition so that the term simplifies to 1
  only when the denominator is nonzero.
The right-hand side of this rule is a large term, so
  exhaustively enumerating all terms from the domain up to its size will
  result in an \egraph with over 3 million terms, far too many
  to be feasible for rule inference.
In \ruler, there is no simple mechanism by which a domain expert can
  narrow the search space in order to learn deeper rules; however,
  \enumo enables users to leverage their domain expertise to find better, deeper
  rules than is possible with exhaustive term enumeration.

Our key goal in this section is to learn conditional versions of unsound rules,
  so let's first find the unsound rule candidates from \autoref{sec:overview-basic}:

\begin{lstlisting}[name=guided]
  # rule candidates over terms up to depth 2
  all_candidates = candidates1.union(candidates2)
  # partition rule candidates using domain-provided rule validator
  (sound, unsound) = all_candidates.partition(|rule| rule.is_valid())
\end{lstlisting}

Now, we can construct a workload that adds a check for
  division by zero:

\pagebreak

\begin{lstlisting}[name=guided]
  guard_wkld = Workload.new()
  guard_pattern = { (if GUARD THEN ELSE) } $\label{ln:guard-pat}$
  for rule in unsound:
    # domain-specific function that returns a workload consisting of terms that
    # appear as the second argument to division
    denominators = rule.denominators() $\label{ln:ext}$
    # construct terms that match the unsound rule candidate, but with a check
    # for division by zero
    guard_wkld = guard_wkld.union(
      guard_pattern
        .plug("GUARD", denominators)
        .plug("THEN", rule.rhs)
        .plug("ELSE", rule.lhs)
    )

\end{lstlisting}

Above, \C{guard_pattern} (line $\ref{ln:guard-pat}$) is a workload consisting of
  a single \sexp, which will serve as a pattern for the workload we are
  constructing.
Note that because \enumo is an embedded DSL, it is possible to encode
  domain-specific extensions simply by writing custom functions.
For example, on line $\ref{ln:ext}$ above, we use a domain-specific function
  to construct a workload.
We loop over the unsound rules, incrementally building the workload
  by using the \C{plug} operator to make a guarded version of each unsound rule.
With this workload in hand, we are ready to learn rules:

\begin{lstlisting}[name=guided]
  candidates = guard_wkld.to_egraph().compress(rules2).find_candidates()
  guard_rules = candidates.minimize(rules2)
\end{lstlisting}

This step closely mirrors the process of learning D2 rules in the
  previous subsection: we convert the workload to an \egraph, compress
  the \egraph using the rules we've already learned, and find candidates.
Finally, we minimize the candidates again using the existing rules.
The final ruleset contains the rules
  \C{(/ a a) \rewritesto \xspace (if a 1 (/ a a))} and
  \C{(/ 0 a) \rewritesto \xspace (if a 0 (/ 0 a))}, both of which are useful, sound
  rewrite rules that avoid unsoundness when the denominator could be zero.
Importantly, we find these rules without enumerating all depth-3
  terms.

\subsection{Learning refined rules with ruleset manipulation}
\label{subsec:ffoverview}

Now, let's consider an alternate approach to candidate generation.
Suppose we want to learn rules for
  transcendental functions (e.g., trigonometric operators).
For the examples in \autoref{sec:overview-basic}
and \autoref{sec:guided-enumeration}, we used cvec matching,
  which requires the user to
  implement an interpreter for the domain.
For transcendental functions,
  equality is undecidable~\citep{boehm}, so
  cvec matching is not possible.
However, these functions \textit{can} be represented
  in terms of other functions
  over rational and complex
  domains\footnote{\url{https://en.wikipedia.org/wiki/Cis_(mathematics)}},
  for which there exist various identities.
For example,
  the functions sine and cosine can
  be represented mathematically by
  $\sin(x) = \frac{\cis(x) - \cis(-x)}{2i}$ and
  $\cos(x) = \frac{\cis(x) + \cis(-x)}{2}$
  where $\cis(x)$ is $e^{ix}$.

Leveraging the compositional nature of \enumo operators,
  we can first synthesize rewrite rules
  over rationals and then
  use them to learn rules for
  trigonometric functions without needing
  to evaluate trigonometric terms directly.
We begin with a set of rewrite rules over rationals
  synthesized previously using an \enumo program.
\pagebreak
\begin{lstlisting}[name=trig]
  initial_rules = Ruleset.from_file("initial.rules")
\end{lstlisting}
Next, we define \emph{exploratory} rules (\autoref{sec:lift})
  that express the
  trigonometric operators in terms of
  the rational and complex operators.
These rules are typically handwritten:
\begin{lstlisting}[name=trig]
  explore = [
    "(sin a) ==> (/ (- (cis a) (cis (- a))) (* 2 I))",
    "(cos a) ==> (/ (+ (cis a) (cis (- a))) 2)",
    "(tan a) ==> (* I (/ (- (cis (- a)) (cis a)) (+ (cis (- a)) (cis a))))"
  ]
\end{lstlisting}
Now, we construct a workload representing trigonometric terms:
\begin{lstlisting}[name=trig]
  consts = { 0 (/ PI 6) (/ PI 4) (/ PI 3) (/ PI 2) PI (* PI 2) }
  wkld = { (OP VAL) }.plug("OP", { sin cos tan }).plug("VAL", consts)
          .filter(Filter.Not(Filter.Contains("(tan (/ PI 2))")))
\end{lstlisting}
This workload represents terms that have
  a single trigonometric operator applied
  to a constant value; notice that we can easily filter out
  \C{(tan (/ PI 2))}, which is undefined.
Finally, we run the \ff algorithm on this workload to discover
  new rules from our initial and exploratory rules.

The \ff algorithm is explained in detail in \autoref{sec:lift}, but at a high level,
it identifies candidates by running known rewrite
  rules and considering merged \eclasses as rule candidates.
By definition, all rules found using this algorithm are derivable from the
  starting ruleset, but the rulesets generated can still be valuable in practice.
In this case, \ff allows us to find rules over the trigonometric operators
  directly, rather than needing to rewrite through large complex terms.

\section{\enumo: A DSL for strategic theory exploration}
\label{sec:slide}

\renewcommand\denote[1]{\ensuremath{\left\llbracket #1 \right\rrbracket}}
\newcommand\set[2]{\ensuremath{\left\{#1 \hspace{3pt} \left\vert \hspace{3pt} #2 \right.\right\}}}
\newcommand\plugsexp{\ensuremath{\mathsf{plug\_sexp}}}

\begin{figure}
  \begin{subfigure}{\linewidth}
    \begin{minipage}[t]{0.45\linewidth}
      \begin{grammar}
      <workload> ::=  Set <s-exp>*
        \alt Union <workload>*
        \alt Filter <filter> <workload>
        \alt Plug <workload> <string> <workload>

      <s-exp> ::=  Atom <string> \alt List <s-exp>+

      \end{grammar}
    \end{minipage}\hfill%
    \begin{minipage}[t]{0.5\linewidth}
      \begin{grammar}

      <filter> ::=
        MetricLt <metric> <\N>
        \alt MetricEq <metric> <\N>
        \alt Contains <pattern>
        \alt Canon <string>+
        \alt And <filter>+ | Or <filter>+ | Not <filter>

      <metric> ::= Atoms | List | Depth
      \end{grammar}%
      \end{minipage}%
    \caption{\enumo workload abstract syntax.}
    \label{fig:workload-grammar}
  \end{subfigure}

  \begin{subfigure}{\linewidth}
    \begin{align*}
      \denote{ \mathsf{Set} ~ ts } &= ts
      &
      \denote{ \mathsf{Filter} ~ \mathit{filter} ~ \W } &=
      \set{t \in \denote{ \W }}{\denote{ \mathit{filter} } (t) = true}
      \\
      \denote{ \mathsf{Union} ~ \W_1 ~ \W_2 } &=
      \denote{ \W_1 }~\cup~\denote{ \W_2 } \hspace{8mm}
      &
      \denote{ \mathsf{Plug} ~ \W_1 ~ \mathit{tgt} ~ \W_2 } &=
        \bigcup_{e \in \denote{\W_1}}
        \denote{ \plugsexp ~ e ~ \mathit{tgt} ~ \W_2 }
    \end{align*}
    \begin{align*}
      \denote{ \plugsexp ~ (Atom ~ s) ~ \mathit{s} ~ \W } &=
      \denote{ \W }
      \\
      \denote{ \plugsexp ~ (Atom ~ s) ~ \mathit{tgt} ~ \W } &=
      \{Atom ~ s\} ~\mathsf{when} ~ s \neq \mathit{tgt}
      \\
      \denote{ \plugsexp ~ (List ~ s_1 ~ \ldots ~ s_n) ~ \mathit{tgt} ~ \W } &=
      \set{List ~ t_1 ~ \ldots ~ t_n}{t_i \in \denote{ \plugsexp ~ s_i ~ \mathit{tgt} ~ \W }}
    \end{align*}
    \caption{\enumo workload semantics.}
    \label{fig:denote-workload}
  \end{subfigure}

  \begin{subfigure}{\linewidth}
    \begin{align*}
      \denote{\mathsf{MetricLt} ~ M ~ n}(t) &= \denote{ M } (t) < n  &
      \denote{\mathsf{Contains} ~ p}(t) &=
        \exists \sigma, \sigma(p) \in \mathsf{subterms}(t) \\
      \denote{\mathsf{MetricEq} ~ M ~ n}(t) &= \denote{ M } (t) = n  &
      \denote{\mathsf{Canon} ~ \vec{a} ~ }(t) &= canon(\vec{a}, t) == t
    \end{align*}
    \caption{
      \enumo filter semantics.
      Metric filters test various measures of term size
        (number of atoms, number of lists, or depth), e.g.,
        \texttt{(+ a b)} has 3 atoms, 1 list, and depth 2.
      The $\mathsf{Contains}$ filter tests whether
        any subterm of $t$ matches a pattern $p$.
      The $\mathsf{Canon}$ filter tests whether
        $t$ is \textit{canonical} with respect to
        a vector of atoms $\vec{a}$, e.g.,
        $\mathsf{Canon} ~ [a, b, c] ~ \mathtt{(+ ~ a ~ b ~ c)}$ is true while
        $\mathsf{Canon} ~ [a, b, c] ~ \mathtt{(+ ~ b ~ a ~ c)}$ is false.
    }\label{fig:denote-filter}
  \end{subfigure}

  \caption{
    Syntax and semantics for the workload fragment of the \enumo DSL.
  }\label{fig:workload}
\end{figure}

This section presents the core of
  the \enumo DSL for
  guided term enumeration and
  incremental rewrite rule inference.
\enumo programs primarily manipulate two kinds of values:
  \textit{workloads} representing sets of terms and
  \textit{rulesets}, which are sets of pairs of patterns.
Both terms within workloads and
  patterns within rewrite rules are
  represented as (untyped) s-expressions.

At a high level, \enumo programs
  typically iterate the following steps:
\begin{enumerate}
  \item Construct a workload $\W$
    representing a search space
    with terms of interest.
  \item Convert $\W$ to an \egraph and
    use the current ruleset to
    merge equivalent \eclasses.
  \item Search this compressed \egraph
    to find candidate rewrite rules,
    i.e., unmerged pairs of \eclasses of terms
    that fuzzing or other techniques suggest
    may be equivalent.
  \item Minimize the candidates:
    remove rules that are redundant
    given the current ruleset.
  \item Add the set of minimized candidates to
    the current ruleset.
\end{enumerate}

\enumo provides a number of operators for
  constructing and manipulating both
  workloads and rulesets, including
  \textit{plugging} and \textit{iterating} workloads
    to build up sets of terms,
  \textit{forcing} to materialize and insert a
    workload of terms into an e-graph,
  \textit{searching} e-graphs built from workloads for
    candidate rewrite rules, and
  \textit{minimizing} rulesets to
    remove redundant rules.
\enumo programs then are just a sequence of bindings
 from variables to workload and ruleset expressions
 embedded in a host language,
 e.g., a simple lambda calculus.

\subsection{Workloads}
\label{subsec:workloads}

\autoref{fig:workload} shows the syntax and semantics of
  workloads in \enumo.
Workloads have four constructors:
  \textsf{Set} represents a literal set of \sexps,
  \textsf{Union} represents unions of workloads,
  \textsf{Filter} represents a subset of terms in a workload, and
  \textsf{Plug} represents substituting
    one workload into another.
Note that workloads only \textbf{represent} sets of terms:
  \C{force} can be applied to a workload to
  materialize the set of
  untyped \sexps (i.e., terms)
  it represents.
This laziness is a key design decision
  that allows \enumo programs to efficiently
  represent and manipulate large sets of terms.

\textsf{Set} and \textsf{Union}
  have straightforward semantics
  (\autoref{fig:denote-workload}).
\textsf{Filter}
 takes a workload $\W$ and a filter predicate $P$,
 and represents the set of terms from $\W$ that satisfy $P$.
\autoref{fig:workload-grammar} shows the
 syntax of filter predicates,
 some of which use term \textit{metrics}.
Metrics count
 the number of atoms, lists, and
 depth of a term.
The semantics for filters is given in \autoref{fig:denote-filter}.
The \textsf{MetricEq} and \textsf{MetricLt} filters
 measure a metric of a term and compare it to a given value.
The \textsf{Contains} filter
 checks whether the given pattern occurs in a term.
The \textsf{Canon} filter
 checks that a given term is canonical with respect
 to a given list of variables.
The \textsf{Not}, \textsf{Or}, and \textsf{And} filters
 are simply the usual logical connectives.

\textsf{Plug}
  allows the user to substitute all combinations
  of terms from one workload in for
  a given variable in another workload.
As the semantics in \autoref{fig:denote-workload} show,
  \textsf{Plug} provides a special kind of substitution
  that performs a Cartesian product:
  \textsf{Plug $\W_1$ $s$ $\W_2$}
  returns a workload
  that denotes a set of
  $\sum_{e \in \denote{\W_1}} |\denote{\W_2}|^{k_e}$ terms\footnote{
    Duplicate terms are removed, so this is an upper bound on the number
    of terms in the resulting workload.
  },
  where $k_e$ is the number of occurrences of $s$ in the term $e$.
The following \enumo snippet demonstrates
  the semantics of \textsf{Plug}:
\begin{lstlisting}
w1 = { X (foo X X) }
w2 = { 1 y }
plugged = w1.plug("X", w2) # { 1 y (foo 1 1) (foo 1 y) (foo y 1) (foo y y)}
\end{lstlisting}

\enumo's operators can be composed into useful,
  reusable strategies beyond the concise reimplementation
  of past work.
As an example,
  \T{iter_metric}, defined in the following \enumo snippet,
  can be used to create size-parameterized workloads:
  \T{iter_metric(W, tgt, Atoms, n)} produces all terms
  from the workload
  with at most $n$ atoms, and \T{iter_metric(W, tgt, Depth, n)}
  produces all terms from the workload up to depth $n$. \T{iter_metric}
  can be used to generate workloads with successively larger terms
  and thus guide the exploration of successively deeper rules
  across domains (\autoref{sec:eval}).

\begin{lstlisting}[
language=Python,
basicstyle=\footnotesize\ttfamily,
numbers=left,
xleftmargin=2.5em]
def iter_plug(W, tgt, n):
  if n <= 0: return W
  return W.plug(tgt, iter_plug(W, tgt, n - 1))

def iter_metric(W, tgt, metric, n):
  return iter_plug(W, tgt, n).filter(metric <= n)
\end{lstlisting}

\paragraph*{Optimizing workloads.}

\textsf{Plug} is the key workload combinator
  for representing search spaces by
  enumerating terms from a grammar.
\textsf{Plug} typically represents
  combinatorially many more terms
  than its arguments,
  but the result of a \textsf{Plug}
  is often \textsf{Filter}ed to
  target a more specific subset of
  the represented terms.
We introduce an essential optimization
  to speed up workload evaluation (forcing)
  that avoids unnecessary work during
  combinatorial substitution by
  pushing some \textsf{Filter}s through \textsf{Plug}s
  according to the following equation:
\begin{align*}
  \denote{\mathsf{Filter} ~ \mathit{filter} ~
          (\mathsf{Plug} ~ \W_1 ~ s ~ \W_2)} &=
  \denote{\mathsf{Filter} ~ \mathit{filter} ~
          (\mathsf{Plug} ~ \W_1 ~ s ~ (\mathsf{Filter} ~ \mathit{filter} ~ \W_2))}
\end{align*}
\noindent
when \textit{filter} is monotonic.
A filter $f$ is \emph{monotonic}
  if, for every term $t$ satisfying $f$,
  every subterm $s \in t$ also satisfies $f$.
Note that the outer \textsf{Filter}
  remains in place even after the optimization,
  as removing it entirely would not
  preserve semantics.
This still yields an exponential reduction
  in the number of terms that must be
  materialized and filtered.
\textit{
  All of the \enumo programs in our evaluation
  depend heavily on this optimization.}

In the current \enumo implementation,
  \textsf{And} and \textsf{MetricLt} filters (among others) are pushed
  through \textsf{Plug}s.
This optimization and its
  monotonicity constraint are inspired by the
  classic relational algebra optimization
  of pushing certain selections through joins~\citep{alice};
  \textsf{Plug}'s combinatorial behavior,
  in some ways, resembles a relational join.

\subsection{\Egraphs and rulesets}
\label{subsec:ruleset-sem}

\renewcommand{\to}{\textrightarrow\xspace}
\newcommand{\head}[1]{\textsf{\sc #1}}
\begin{figure}
  \footnotesize
  \begin{minipage}[t]{0.5\linewidth}
  \centering
    \raggedright\tt
    \head{E-graph Generating Operators} \\
    \vspace{4pt}
    \toeggl  : $\T{workload}$ \to \eggl \\
    $\T{eqsat}$    : \eggl \to \ruleset \to \eggl \\
    $\T{compress}$ : \eggl \to \ruleset \to \eggl \\[1mm]
    \vspace{4pt}
    \head{Rule Testing Operators} \\
    \vspace{4pt}
    $\T{is\_saturating}$ : \rl \to $\T{bool}$ \\
    $\T{is\_valid}$      : \rl \to $\T{bool}$ \\
    $\T{can\_derive}$    : \ruleset \to \rl \to $\T{bool}$
    \vspace{4pt}
  \end{minipage}
  \begin{minipage}[t]{0.65\linewidth}
  \centering
    \raggedright\tt
    \head{Ruleset Generating Operators} \\
    \vspace{4pt}
    $\T{find\_candidates}$    : \eggl \to \ruleset \\
    $\T{partition}$          : \ruleset \to (\rl \to $\T{bool}$) \to \ruleset $\ast$ \ruleset \\
    $\T{minimize}$           : \ruleset \to \ruleset \to \ruleset \\
    $\T{union}$              : \ruleset \to \ruleset \to \ruleset \\
    $\T{candidates\_by\_diff}$ : \eggl \to \eggl \to \ruleset
  \end{minipage}
  \caption{
    \enumo's operators over rulesets and \egraphs.
  }\label{fig:ruleops}
\end{figure}

In addition to novel, programmable term enumeration,
 \enumo also provides primitives to
 create and manipulate \egraphs and rulesets.
\autoref{fig:ruleops} shows
  these operators and their types.
Many mirror parts of
  earlier monolithic theory explorers;
\enumo's key insight lies in turning such tools
  ``inside out'' to provide their components
  as composable operators in a DSL that
  allows users to strategically guide
  the search for rewrites and
  incrementally build rulesets.

\paragraph*{\Egraph operators.}
For the purposes of the \enumo language definition,
  an \egraph is an abstract data type
  that provides the operations described
  in \autoref{sec:bg-egraphs}.
A typical \enumo program (\autoref{sec:overview})
 converts a workload into an \egraph
 using the \toeggl operator
 before proceeding to candidate generation.
The resulting \egraph represents every term in the set
 denoted by the workload.

The \texttt{eqsat} and \texttt{compress} operators
 both run equality saturation on the given \egraph
 with the given ruleset.
The former allows the \egraph to grow while the latter
  only applies merges over existing terms in the \egraph.
From \enumo's perspective,
 \texttt{compress}'s main purpose is to
 remove redundancy from the \egraph implied
 by a ruleset of already-learned rewrites.
Because we build on the \egg library,
 \texttt{eqsat} in our implementation
 is also parameterized over a \emph{strategy}
 that determines how a ruleset is applied to the \egraph
 (\autoref{subsec:cmp}).

\paragraph*{Ruleset operators.}

A ruleset is a set of rewrite rules,
 where each rule is a pair of patterns.
Rulesets can be
 read from or written to a file,
 manipulated using the ruleset operators in \autoref{fig:ruleops},
 and used to perform equality saturation on \egraphs
 using the \texttt{eqsat} operator described above.

In a typical \enumo program,
 the \T{find\_candidates} operator
 is used to infer a ruleset from an \egraph.
\T{find\_candidates}
 is parameterized on a user-provided interpreter
 which is used to
 identify likely sound rule candidates
 by evaluating the terms over a set of inputs (fuzzing);
 terms that disagree are certainly not equivalent,
 but those that agree may be~\citep{ruler,bansal}.
\renewcommand{\eval}{\textsf{eval}}
\newcommand{\repr}{\textsf{repr}}
To define \T{find\_candidates} formally,
 let $\repr(e)$ denote a \textit{representative} term from
 \eclass $e$
 and let $\eval(t)$ denote the result of evaluating $t$
 over a set of input values.
Then \T{find\_candidates}
  on \egraph $E$ returns a set of rules:
\[
  \set{\repr(e_l) \rewritesto \repr(e_r)}{
    \text{$e_l, e_r \in E$}.\
    \eval(\repr(e_l)) = \eval(\repr(e_r))
  }
\]

The \T{partition} operator takes a \ruleset $R$
  and a predicate $P$ over rules and
  returns $(R_1, R_2)$ such that
  $R = R_1 \cup R_2$,
  $r \in R_1 \rightarrow P(r)$, and
  $r \in R_2 \rightarrow \neg P(r)$.
\autoref{fig:ruleops}
 contains two such predicates, \T{is\_valid} and \T{is\_saturating}.
The \T{is\_valid} predicate
 checks whether a rule $\ell \rewritesto r$
 is valid for all inputs,
 using a user-provided verifier
 for the domain.
Depending on the domain,
 the verifier may use SMT, model checking,
 or other techniques like fuzzing.
The built-in \T{is\_saturating} predicate
 checks whether a rule is \textit{saturating},
 i.e., applying the rule to an \egraph
 will not increase the size of the \egraph,
 measured as the number of \eclasses.
At a high level,
  saturating rules are those whose right-hand side pattern
  only contains subterms that appear in the left-hand side,
  except potentially for the root operator.
For example,
  $x + y \rewritesto y + x$ is saturating
  since all non-root subterms in the right-hand side ($x$ and $y$) also
  occur in the left-hand side but
  $x + (y + z) \rewritesto (x + y) + z$ is not
  since the right-hand side contains a non-root subterm $x + y$
  that does not appear in the left-hand side.
Applying only saturating rules
  to an \egraph
  is guaranteed to reach a fixpoint
  past which further application of the rules
  will no longer change the \egraph.

The \T{can_derive} operator
 tests whether a \ruleset $R$
 can derive a rule $\ell \rewritesto r$,
 discussed below in \autoref{subsec:derivability}.
The \T{minimize} operator
 takes a ruleset $R$ and prior rules $P$
 and \textit{minimizes} $R$ with respect to $P$,
 guaranteeing that $
  \mathtt{minimize}(R, P) = R' \implies
  \forall r \in R \setminus R',
  \T{can\_derive}(P \cup R', r)
 $.
Conceptually,
 this filters $R$ to a small subset $R'$ of rules
 such that $r \in R' \rightarrow \neg \T{can\_derive}(P, r)$.
However, \enumo's \T{minimize} operator provides
 an optimization that batches these checks to simultaneously
 eliminate redundant rules, initially described in \citet{ruler}.

The final core operator
  is \T{candidates\_by\_diff},
  which takes two \egraphs $e_1$ and $e_2$ and
  returns an inferred ruleset.
\T{candidates\_by\_diff}
  infers candidate rewrite rules from
  \eclasses which merged during an \eqsat run,
  i.e., terms which a given ruleset could prove equivalent.
Typically, $e_2$ is the result of
  running \eqsat on $e_1$ with ruleset $R$.
If, by application of $R$, the equivalence between
  terms $t$ and $t'$ is discovered,
  then \T{candidates\_by\_diff} will learn a
  rule candidate by extracting the best expression from the \eclasses
  representing $t$ and $t'$ in $e_1$.
\T{candidates\_by\_diff} enables rule synthesis for new domains
  which prior work could not support. This utility is
  briefly exemplified in~\autoref{subsec:ffoverview},
  and formally presented in~\autoref{sec:lift}.

\subsection{Discussion on derivability}
\label{subsec:derivability}

Given two rulesets, how do we know which is better?
While it may be tempting to use ruleset size as a proxy
  for ruleset quality, more rules are not necessarily better
  because overly redundant rules lead to slower performance of \eqsat
  systems.
A small set of simple rules is often easier to maintain and
  debug than a large set of complicated rules.
On the other hand, a ruleset with too few rules is less useful
  because fewer equivalences will be found,
  especially since resource limits restrict
  the number of iterations of \eqsat.
Since saturation is rare in practice, it is often helpful to have
  \textit{some} redundancy in the rulesets to get better results
  under given resource limits (see \autoref{sec:lift}).
 A ruleset's \textit{proving power} under given resource
  limits is subtle and difficult to estimate.
In this section, we define ruleset \textit{derivability}, a
  metric for measuring proving power that we use for comparing rulesets.

\paragraph*{Derivability.}
Prior work has not established a standard definition of derivability
  in the context of \eqsat.
In this paper, we formalize two ``obvious'' definitions of derivability:
  \lhsandrhs and \lhs.
To test whether a ruleset $R$ can derive a rule $\ell \rewritesto r$
  under given resource limits, we use the \eqsat procedure
  (\autoref{fig:eqsat-alg}).
The function \C{timeout}
  determines when to stop the \eqsat loop based on available resources
  (e.g., node count, iteration, or time bounds).
If running \eqsat using ruleset $R$ causes \eclasses representing
  $\ell$ and $r$ to merge, we say $\ell \rewritesto r$ is \textit{derivable}
  from $R$ under given resource bounds.
The \lhsandrhs derivability metric measures whether the equivalence between
  $\ell$ and $r$ can be recovered by applying the rules in $R$ to
  an \egraph initialized with both $\ell$ and $r$.
In contrast, the stronger \lhs definition for derivability states
  that $\ell \rewritesto r$ can be recovered given only $\ell$.
Prior work used \lhsandrhs derivability~\citep{ruler}.

In the context of \eqsat, the initial state of the \egraph interacts with
  resource limits in subtle ways because it changes what terms are
  available during e-matching.
Rules in $R$ must find concrete terms in the \egraph that match the
  left side of the rule in order to add the right side and merge the two
  \eclasses.
Changing the initialization of the \egraph changes what rule matches
  are possible.

To illustrate the difference between \lhs and \lhsandrhs derivability,
  consider the rule $a \rewritesto b$
  (where $a$ and $b$ are arbitrary patterns) and a ruleset containing
  the rule $b \rewritesto a$.
In an \egraph initialized with both $a$ and $b$ (\lhsandrhs), the rule
  $b \rewritesto a$ will fire and the \eclasses will merge, so the
  rule $a \rewritesto b$ will be considered derivable.
In an \egraph initialized with just $a$ (\lhs), $b \rewritesto a$
  will not fire, so the rule $a \rewritesto b$ will not be considered
  derivable.
\lhs and \lhsandrhs derivabilities may also require different resource
  limits.
For example, consider using
  $R = \{a \rewritesto b, b \rewritesto c, c \rewritesto b \}$ to
  derive the rule $a \rewritesto c$.
Under \lhsandrhs, the \eclasses representing $a$ and $c$ will
  merge within a single iteration of \eqsat.
In contrast, using \lhs derivability, recovering the equality between $a$ and
  $c$ will take two iterations of \eqsat.
First, the rule $a \rewritesto b$ will fire, creating an \eclass for
  $b$ and merging it with $a$'s \eclass.
In the second iteration, the rule $b \rewritesto c$ will fire, creating
  an \eclass for $c$ and merging it with the \eclass that represents
  $a$ and $b$, thus recovering the equivalence between $a$ and $c$.
This example shows that \lhsandrhs derivability may be able to derive equivalences
  in fewer iterations (i.e., using fewer resources) than \lhs because it can
  match on the left-hand and right-hand sides simultaneously.

In general, \lhs derivability is more conservative.
Anecdotally, we find that it is preferable
  when the user is interested in optimization-based
  \eqsat applications where an \egraph is initialized
  with a single term $t$, and \eqsat is used to find
  a better, equivalent version of $t$.
In contrast, \lhsandrhs derivability is looser,
  but may be appropriate in equivalence-checking
  \eqsat applications where two terms $t_1$ and $t_2$
  are added to an \egraph and \eqsat is used only to
  determine whether their \eclasses merge.

\section{A fast-forwarding theory explorer}
\label{sec:lift}
\label{subsec:ff-nointerp}
\label{subsec:perf}

This section presents a new
  \textit{fast-forwarding} theory exploration
  algorithm, which has two key applications.
First, as \autoref{subsec:ffoverview} showed,
  it enables rewrite rule inference
  for domains where writing an interpreter
  is prohibitively difficult.
Second,
  it mitigates performance problems that arise
  when resource limits interact poorly with
  the kinds of rules in a ruleset.

The \textit{kinds} of rules that comprise a ruleset
  significantly affect performance,
  even in efficient \eqsat-driven systems.
Since reaching saturation
  in an \egraph is rare in practice,
  iteration and/or node limits are used
  to ensure termination.
As a result,
  two rulesets can have vastly different
  performance
  even if they are equivalent under derivability.

The key motivation behind
  the fast-forwarding algorithm is that
  the ``right'' set of rules
  can help \textit{fast-forward} equality saturation by
  skipping intermediate derivations.
Skipping intermediate derivations has two benefits:
  first, it requires fewer iterations to prove
  a target equivalence;
  second, it often
  reduces the number of intermediate terms in the \egraph,
  reducing unhelpful rewriting on these terms.
Determining the ``right'' rules requires domain knowledge
  and depends on the application.
To that end, we assume that the user can provide
  a set of \textit{allowed} ($\mathcal{A}$) and
  \textit{forbidden} ($\mathcal{F}$) operators.
We then say that
  if a pattern $p$ contains \textit{any} operator $o \in \mathcal{F}$,
  then $p$ is forbidden.
If \textit{all} operators in $p$ are allowed,
  then the pattern is allowed.
Since a rule is simply a pair
  of patterns,
  these definitions extend to rules.

For example, in the case of the trigonometric
  rule synthesis task in \autoref{subsec:ffoverview},
  the allowed operators are \texttt{sin}, \texttt{cos}, \texttt{tan},
  \texttt{PI}, \texttt{+, -, $\times$}, etc., and
  the forbidden operators are
  \texttt{cis} and \texttt{I} because we wanted \enumo to synthesize
  rewrite rules over the trigonometric domain only.
Recall that this task also required an additional set of
  \textit{exploratory} ($\mathcal{E}$) rewrite rules that relate
  terms with allowed operators to terms with \emph{other} operators.
Crucially, these ``other'' operators can be either allowed or
  forbidden.
The intuition behind $\mathcal{E}$
  is that it helps \textit{explore}
  new equivalences between allowed terms in the \egraph by
  applying a \textit{known} set of rewrites over terms containing the
  \emph{other} operators
  (shown by \texttt{explore} in \autoref{subsec:ffoverview}).

\begin{figure}
\begin{lstlisting}
def fast_forward_naive($\mathcal{W},\, \mathcal{R}$):
  $\mathcal{G}$ = $\mathcal{W}$.to_egraph() # convert workload to e-graph
  $\mathcal{G'}$ = $\mathcal{G}$.compress($\mathcal{R}$) # run equality saturation with all rules
  # any two terms from the unions are potential candidates:
  candidates = candidates_by_diff($\mathcal{G}$, $\mathcal{G'}$)
  return candidates.minimize($\mathcal{R}$) # minimize candidates
\end{lstlisting}
  \caption{A naive algorithm for fast-forwarding theory exploration which
  applies \eqsat to terms represented by $\mathcal{W}$ using
  \T{compress}.}
\label{fig:ff-naive}
\end{figure}

\autoref{fig:ff-naive} shows a naive implementation
  of fast-forwarding
  using the set of core \enumo operators
  from \autoref{sec:slide}.
The process consists of
  applying \T{eqsat} to a workload representing
  \textit{allowed} terms
  using a ruleset, $\mathcal{R}$, which contains both
  allowed and forbidden rules.
First, it creates an \eggl from the terms obtained by
  evaluating the workload.
Then, it \textit{shrinks} the \egraph using the
  \T{compress} operator.
\T{compress} is an \eqsat \textit{strategy} (\autoref{fig:ruleops}) that prevents
  the \egraph from getting intractably large.
It applies the rewrites on a duplicate of the original
  \egraph and only copies the
  equivalences back.
It adds no new \enodes or \eclasses to the original \egraph.
The next step in the algorithm extracts candidates from $\mathcal{G}$
  based on the equalities discovered in $\mathcal{G'}$
  using a cost function that penalizes forbidden operators.
Finally, it minimizes the resulting ruleset as explained
  in \autoref{sec:slide}.
Notice that this naive algorithm
  simply performs a single phase of \T{compress} with
  \textit{all} the rules.

\begin{figure}
\begin{lstlisting}
def fast_forward($\mathcal{W},\, \mathcal{R},\, \mathcal{E}$):
  $\mathcal{G}$ = $\mathcal{W}$.to_egraph() $\label{ln:5}$ # convert workload to egraph
  allowed = $\{r ~ \in ~ \mathcal{R} ~ | ~ \forall ~ o ~\in~ (ops(r.lhs) ~ \cup ~ ops(r.rhs)), ~ o ~\in~ \mathcal{A} \}$  $\label{ln:6}$
  $\mathcal{G'}$ = $\mathcal{G}$.compress(allowed) $\label{ln:7}$ # compress the egraph with allowed rules
  $\mathcal{G''}$ = $\mathcal{G'}$.eqsat($\mathcal{E}$) $\label{ln:11}$ # grow the egraph with exploratory rules
  $\mathcal{G'''}$ = $\mathcal{G''}$.compress($\mathcal{R}$) $\label{ln:12}$ # compress to find equalities with all of $\mathcal{R}$
  candidates = candidates_by_diff($\mathcal{G'}$, $\mathcal{G'''}$) $\,$# extract learned rules with no ops in $\mathcal{F}$
  return candidates.minimize(allowed) # minimize candidates
\end{lstlisting}
\caption{A practical fast-forwarding theory exploration algorithm that
  approximates the naive version. \textit{ops} is a helper function that returns
  all the operators in a term.}
\label{fig:ff-actual}
\end{figure}

\paragraph*{A practical algorithm.}
Unfortunately, the naive algorithm in \autoref{fig:ff-naive}
  does not find useful
  rules in practice for two reasons.
First, it does not scale:
  for large workloads, the \egraph blows up and
  resource limits (e.g., timeout, iterations) are exhausted
  before useful equivalences emerge.
Second, exploring in a breadth-first manner
  prevents finding interesting fast-forwarding opportunities,
  which only occur after several rounds of \eqsat.

Instead, we propose a more practical, approximate algorithm
  that applies \eqsat
  in a more strategic way by leveraging a user's domain
  knowledge in the form of $\mathcal{E}$,
  $\mathcal{F}$, and $\mathcal{A}$.
The algorithm in \autoref{fig:ff-actual}
  \textit{selectively} grows and compresses
  the \egraph using the rules provided by the user.
The algorithm first creates an \eggl
  from the terms represented by $\mathcal{W}$,
  then compresses the \egraph with allowed rules
  (lines \ref{ln:6}--\ref{ln:7}).
This step  \textit{shrinks} the \egraph with known
  equivalences.
In the next step,
  the algorithm \textit{grows} the \egraph with
  $\mathcal{E}$ (line \ref{ln:11}).
Crucially, this step does \emph{not} use \T{compress};
  it performs simple \T{eqsat} that introduces
  new terms and equivalences in the \egraph.
The final \eqsat step applies another round of compression
  using all the rules in $\mathcal{R}$, discovering new
  fast-forwarded rules.
The minimization step in this algorithm
  uses the \T{allowed} rules instead of the entire
  ruleset in order to avoid forbidden operators
  in the minimized ruleset.
The rest of the algorithm is similar to \autoref{fig:ff-naive}.

\subsection{Comparing different scheduling strategies}
\label{subsec:cmp}
The key idea in our fast-forwarding algorithm is to
  perform \eqsat in phases,
  using subsets of $\mathcal{R}$ to selectively
  grow and compress the \egraph.
To understand how this affects performance,
  we ran an experiment to evaluate and compare
  the difference between using
  \C{eqsat} and \C{compress}
  in \autoref{fig:ff-actual}.
We used a workload of 287 terms
  from the domain of trigonometric operators
  ($\sin$, $\cos$, $\tan$, $\pi$, $\pi/2$, etc.).
\autoref{table:ff-cmp} shows the
  results of the comparison.
The first two rows, which use \C{eqsat} in all three phases, do not terminate
  within 20 minutes.
These treatments demonstrate the importance of \C{compress}, which does not
  allow the \egraph to grow.
The next two rows use \C{compress} in all three phases.
The third row does not split up the rules in $\mathcal{R}$ and simply runs
  \C{compress($\mathcal{R}$)} three times.
The fourth row compresses with the allowed rules ($\mathcal{A}$) in
  Phase 1, the exploratory rules ($\mathcal{E}$)
  in Phase 2, and all rules ($\mathcal{R}$) in Phase 3.
Both of these treatments finish within seconds, but do not find any new rules.
The approach in the last row, which corresponds to the actual \ff algorithm
  described in \autoref{fig:ff-actual}, finds 4 useful trigonometric
  identities in about 3 minutes.
This experiment demonstrates the importance of using \C{eqsat} and \C{compress}
  together to strategically grow and compress the \egraph.

\begin{table}[h]
  \footnotesize
\begin{tabular}{llllc}
  Phase 1 & Phase 2 & Phase 3 & Time (s) & \# Rules with Trig Operators \\ \cline{1-5}
eqsat ($\mathcal{R}$) & eqsat ($\mathcal{R}$) & eqsat ($\mathcal{R}$) & Timeout & - \\
eqsat ($\mathcal{A}$) & eqsat ($\mathcal{E}$) & eqsat ($\mathcal{R}$) & Timeout & - \\
compress ($\mathcal{R}$) & compress ($\mathcal{R}$) & compress ($\mathcal{R}$) & 18.66 & 0 \\
compress ($\mathcal{A}$) & compress ($\mathcal{E}$) & compress ($\mathcal{R}$) & 9.42 & 0 \\
compress ($\mathcal{A}$) & eqsat ($\mathcal{E}$) & compress ($\mathcal{R}$) & 175.78 & 4 \\
\end{tabular}%
\caption{Comparing \texttt{compress} and \texttt{eqsat} with different
    subsets of $\mathcal{R}$ for the three phases of \autoref{fig:ff-actual}.
    $\mathcal{A}$ is the allowed rules of $\mathcal{R}$ and
    $\mathcal{E}$ is the exploratory rules of $\mathcal{R}$.
    The last row corresponds to \autoref{fig:ff-actual};
    it is the only treatment that finds any trigonometric rules.}
\label{table:ff-cmp}
\end{table}

\section{Evaluation and case studies}
\label{sec:eval}

\paragraph*{Implementation.}
\enumo is an embedded DSL,
  implemented as a \rust library.
The entire implementation
  is 3095 LOC, including unit tests but excluding
  the implementations of the various domains.
The domains together add another 4430 LOC,
 which comprises a grammar, evaluator, and validator for
 each basic domain,
 and a grammar and fast-forwarding rules for
 each domain employing fast-forwarding.
The various \enumo programs add up to 718 LOC.
Our \enumo implementation, the domain implementations, and all the \enumo programs
  are publicly available.

To evaluate our contributions,
  this section answers the following
  research questions.

\begin{enumerate}

\item How does guided enumeration in \enumo
  compare to prior work on rewrite rule synthesis?
  (\autoref{subsec:eval-enumo})

\item Can \enumo scale to larger grammars than
  existing tools can handle?
  (\autoref{subsec:eval-enumo})

\item Can \enumo's fast-forwarding algorithm enable rule
  inference for new domains that prior work could not support?
  (\autoref{subsec:eval-ff})

\item How does fast-forwarding impact client applications
    in terms of performance and results?
    (\autoref{subsec:eval-ff})

\item Can \enumo be extended with new synthesis techniques,
    and do the new operators compose with existing ones?
    (\autoref{sec:case-study-llm})
\item Do the abstractions in \enumo
    enable cross-domain rule synthesis?
    (\autoref{sec:case-study-bv})

\end{enumerate}

\subsection{Guided search with \enumo}
\label{subsec:eval-enumo}

To evaluate \enumo's guided search,
  we conducted the following experiments on a
  64-bit Linux machine with 32 GB RAM,
  running Ubuntu 22.04.2 LTS.

\begin{table}[h]
\resizebox{\textwidth}{!}{%
\begin{tabular}{lcllcc}
Domain   & \enumo LOC  & \# \enumo (Time)   & \# \ruler (Time)  & \enumo $\rightarrow$ \ruler (Time) & \ruler $\rightarrow$ \enumo (Time)  \\ \cline{1-6}
bool & 41 & 64 (0.35)   & 51 (0.05) & 100\% (0.01), 100\% (0.01) & 87.5\% (5.29), 96.9\% (0.01) \\
bv4  & 19 & 180 (7.13)  & 84 (0.96) & 100\% (0.17), 100\% (0.03) & 38.3\% (3.67), 41.1\% (4.32) \\
bv32 & 18 & 120 (48.78) & 78 (13.1) & 100\% (0.15), 100\% (0.01)  & 58.3\% (1.41), 60.0\% (2.08) \\
rational & 57 & 123 (6.34) & 113 (97.9) & 97.3\% (0.73), 100\% (0.09) & 52.0\% (18.66), 58.5\% (22.24) \\
\end{tabular}%
}
\caption{Results comparing \enumo to \ruler.
  $ R_1 ~ \rightarrow ~ R_2$ indicates using $R_1$ to derive
  $R_2$ rules.
  We report both \lhs and \lhsandrhs derivability (in that order), separated by commas.
  The numbers in parentheses are times in seconds.
}
\label{table:oopsla}
\end{table}

\subsubsection{Comparing rulesets with prior work}
\label{subsubsec:e-vs-r}
\ruler~\citep{ruler} is a state-of-the-art
  tool for automatically
  synthesizing rewrite rules targeted towards \eqsat-driven
  systems.
We compare the rules generated by \enumo
  against those generated by \ruler\footnote{bool, bv4, and bv32
  rules are taken from the \ruler artifact. The artifact did not contain
  rational rules, so those were generated following the instructions in
  the \ruler repository on GitHub.},
  finding that
  rulesets from small \enumo programs outperform
  those from \ruler,
  which uses a hard-coded rule-finding
  strategy.
We wrote \enumo programs for
  each of the domains showcased in \ruler:
  \texttt{bool}, \texttt{bv4}, \texttt{bv32}, and \texttt{rational}.
These programs call \T{recursive_rules},
an \enumo-provided utility function (\autoref{fig:rec-rules}).
From a user-provided grammar $\mathcal{G}$ that specifies literal terms,
  unary operators, and binary operators, \T{recursive_rules} builds
  workloads of increasing size, then finds and validates rules from
  the workloads, using rules it finds along the way to avoid redundancy
  in the final ruleset.
This algorithm
 replicates \ruler's core loop in just a few lines of
\enumo code, and could easily be written by an \enumo user,
highlighting the flexibility and power of the tool.
\begin{figure}
\begin{lstlisting}
lang = { LIT (UOP EXPR) (BOP EXPR EXPR) }
def recursive_rules($\mathcal{G}$, metric, n):
  if n == 0: return []
  rec_rules = recursive_rules($\mathcal{G}$, metric, n - 1)
  workload = iter_metric(lang, "EXPR", metric, n)
               .plug("LIT", $\mathcal{G}$.lits).plug("UOP", $\mathcal{G}$.uops).plug("BOP", $\mathcal{G}$.bops)
  rules_n = workload
              .to_egraph()
              .compress(rec_rules)
              .find_candidates()
              .minimize(rec_rules)
  return rec_rules.union(rules_n)
\end{lstlisting}
\caption{The \T{recursive_rules} function. $\mathcal{G}$ is a struct that
  specifies a grammar, containing workloads for literal expressions,
  unary operators, and binary operators. \T{metric} is the \enumo metric
  used to define an upper bound on terms, and \T{n} is the size limit.
  \T{recursive_rules} incrementally builds a ruleset by learning rules
  over terms from the domain of increasing size up to the
  specified limit.}
\label{fig:rec-rules}
\end{figure}

After running our \enumo programs,
we compared the derivability of the generated rulesets
to those produced by \ruler,
using the same grammar and interpreter.
We found that for rational arithmetic,
  \ruler learns rules over division by assuming that
  the denominator is not zero\footnote{To mitigate the resulting
  unsoundness, \ruler used a custom rule application
  strategy from the \egg~\citep{egg} library.}.
We removed this unsound
  assumption from \ruler and re-synthesized its
  rational arithmetic rules for our comparison.
We also found that for \C{bool}, \C{bv4}, and \C{bv32}, \ruler's hard-coded
  term enumeration loop does not enumerate any constant values.
As a result, \ruler fails to learn simple identities like \C{(\& ?a (~ ?a)) \rewritesto \, false}
  and \C{(<< ?a 0) \rewritesboth \, ?a}.
Failure to enumerate these constants explains why \ruler is unable to derive
  many of \enumo's rules for these domains.
We adapted the \enumo programs to not enumerate constants, and we find that
  \ruler can derive 100\% of the \C{bool} rules, 90.7\% of the \C{bv4} rules,
  and 87.8\% of the \C{bv32} rules\footnote{Both \lhs and \lhsandrhs derivability
  result in these percentages.}.

We found that \enumo rulesets were able to derive
  all of \ruler's rules using \ruler's
  own \lhsandrhs derivability metric
  (\autoref{subsec:derivability}, \autoref{table:oopsla}).
The reverse is not true.
Using the more conservative \lhs metric, \enumo rulesets derive
  a higher percentage of \ruler's rules than \ruler's rules can derive of
  \enumo's.
Both measures suggest that \enumo rulesets have greater proving power
  than their \ruler counterparts.
This result demonstrates that small (\C{<}60 LOC), simple
\enumo programs outperform \ruler while also providing users
with increased flexibility and transparency.

\subsubsection{Scaling to large grammars: A Halide case study}
\label{subsec:halideeval}
Halide~\citep{halide} is a programming language
  for high-performance
  image processing.
A major component of the Halide compiler is
  a traditional term rewriting
  system~\citep{julie-halide} that performs optimizing
  program transformations using a set of handwritten rules.
Halide has a large grammar, totaling 17 operators,
  which include boolean,
  arithmetic, and
  comparison operators.
Halide does not use an \eqsat engine
  for applying the rewrite rules.
Nevertheless, inspired by the domain,
  particularly due to the size of its grammar,
  we developed an \enumo program to evaluate
  the scalability of \enumo's workload-guided strategy.

Halide's handwritten ruleset was collected by scraping
source files from the latest commit in the Halide
  repository\footnote{\url{https://github.com/halide/Halide/commit/e7f78600e10956b44e8f214c686f310211b0d836}}
  and removing rules that we could not parse as \enumo rules.
These included rules with side conditions, rules with unsupported operators,
  and rules with unbound variables on the right-hand side of the rule.
After this process, we were left with 725 rules.

To see how prior work~\citep{ruler}
  would perform on a large domain,
  we implemented the Halide grammar in \ruler. \ruler's
  implementation was able to run for
  just one iteration (further iterations did not terminate),
  synthesizing a total of 90 rules in about 3 seconds.
  This ruleset was able to derive
  only 18 of the 725 original rules (2.5\%)
   using both derivability metrics (\lhs, \lhsandrhs).

Without leveraging guided search, \enumo scales similarly to Ruler
  for exhaustive exponential searches.
A simple \enumo program that exhaustively enumerates Halide terms up to size
  5 is able to derive 327 of Halide's 725 rules.
The exhaustive \enumo program only outperforms \ruler because \ruler
  enumerates by depth, which grows much faster than size.
An \enumo program that enumerates by depth times out after depth 2 and learns
  rules that can only derive 13 of Halide's rules, which is similar to
  \ruler's behavior.

However, the key benefit of \enumo's guided enumeration is decoupling
  the grammar and the workload size.
In \ruler, terms are enumerated exhaustively from the grammar up to a
  certain size, so with a larger grammar, \ruler hits resource limits
  faster.
In contrast, term enumeration in \enumo is separate from the
  grammar itself, allowing users to define workloads that represent
  different subsets of the search space.
With \enumo's operators, it is easy to compose workloads together,
  enabling a piecewise rather than total approach to term enumeration.
This composability makes it possible to synthesize rulesets that are
  larger and deeper than would be possible with a one-shot theory
  exploration tool like \ruler.

To evaluate whether \enumo's
  guided search would help to find deeper, more complex Halide rules,
  we wrote a 141-line \enumo program
  leveraging both exhaustive and custom enumeration.
  First, we exhaustively enumerated terms over subsets of Halide's
  operators---boolean, arithmetic,
  comparison\footnote{For both \ruler and \enumo,
  we enumerated terms over
  15 of the 17 operators, skipping $/$ and $\implies$ since
  most of those rules had side conditions.}%
  ---up to 5 atoms, beyond which point this strategy
  is computationally infeasible.
  We then enumerated
  terms with \textit{all} of
  Halide's operators up to
  4 atoms in size.
Finally, we created custom
  workloads guided by domain knowledge,
  selectively generating terms too large to be
  found using the exhaustive approach,
  such as \C{(select a (min b c) (max d c))}.
  These workloads leverage \enumo features
  such as canonicalization
  (\autoref{sec:slide}), which eliminates many duplicate terms,
  reducing one workload
  from 52,491 to 9,233 terms---an 82\%
  decrease\footnote{More details on how the Halide programs
  were written are available in \autoref{subsec:halide-appendix}.}.
 Ultimately, our \enumo program produced a ruleset of 840 rules
  capable of deriving
  84.0\% and 90.9\% of the handwritten ruleset using
  the \lhs and \lhsandrhs derivability
  metrics, respectively.

As mentioned in \autoref{subsec:derivability}, larger rulesets are not
  necessarily better than smaller rulesets.
In this case study, however, the smaller ruleset (from one iteration of \ruler)
  has measurably less proving power than the larger ruleset
  (from the \enumo program), so the additional rules are justified.
Synthesizing rulesets for large grammars is not feasible
  with tools that rely on exhaustive term enumeration,
  but with \enumo, it is possible to build
  rulesets incrementally, so grammar size is not a
  limiting factor.
This section shows that a small program
using \enumo's novel guided search finds better rules than
the state of the art can.

\subsection{Fast-forwarding}
\label{subsec:eval-ff}

In this section,
  we evaluate our new \ff algorithm (\autoref{sec:lift})
  in two new domains to learn rewrite rules
  that other state-of-the-art tools
  do not support.
We also show that synthesized rules from \enumo
  can be easily integrated with
  existing \eqsat-based synthesis tools.

\begin{table}[h]
\resizebox{\textwidth}{!}{%
\begin{tabular}{lllccc}
Domain      & \enumo LOC  & \# \enumo (Time)  & \# \herbie  & \enumo $\rightarrow$ \herbie (Time) & \herbie $\rightarrow$ \enumo (Time)  \\ \cline{1-6}
  Exponential &  186  & 40 (3.58)    & 82 &  39.0\% (1.03), 43.9\% (1.25)   & 100\% (0.24), 100\% (0.01) \\
  Rational    &   82  & 129 (276.16) & 87 &  70.1\% (54.77), 73.6\% (55.19) &  -, -\\
  Trigonometric  &  160  & 56 (631.94)  & 45 &  33.3\% (4.20), 35.6\% (7.36) & 46.4\% (0.84), 46.4\% (2.27)\\
\end{tabular}%
}
  \caption{
  Derivability comparison between rules from \enumo and \herbie.
  As in \autoref{table:oopsla},
    $R_1 ~ \rightarrow ~ R_2$ indicates using
    $R_1$ to derive $R_2$ rules.
  For Exponential and Trigonometric, we include the trusted arithmetic
    rules defined below (\C{RAT}, and \C{R} without its division
    rules) when computing the derivability.
  We report both \lhs and \lhsandrhs derivability
    (in that order), separated by commas. The numbers in parentheses
    are times in seconds.
 ``-'' indicates that the derivability test
    could not be completed due to \herbie's
    unsound rules (\autoref{para:herbie}).
  We integrate these rules for
    end-to-end runs of \herbie~\citep{herbie} and
    Megalibm~\citep{megalibm} (\autoref{subsubsec:numbers}).}
\label{table:herbie}
\end{table}

\subsubsection{Numeric domain}
\label{subsubsec:numbers}

We used the fast-forwarding algorithm together with
  guided enumeration to infer rewrite rules
  for the domain of transcendental functions.
To evaluate the quality of the rulesets,
  we integrated the rules into two existing
  rewrite-rule-based synthesis tools,
  \herbie~\citep{herbie} and
  Megalibm~\citep{megalibm}.

\paragraph*{Trigonometric and exponential representation.}
\label{eval:trig-rep}

Recall that sine and cosine have representations
  in terms of the complex exponential $\cis(x)$
  (\autoref{subsec:ffoverview}).
The trusted rules for the Trigonometric domain fall into three groups, which we refer
  to throughout the evaluation:
\begin{itemize}
  \item \C{R}: 57 automatically generated arithmetic rules,
    including 13 rules involving division that are sound only when their divisors are nonzero\footnote{
      In our \enumo programs for this domain, we are careful to exclude undefined terms from the \egraph,
      so the potentially unsound rules are harmless. Herbie uses unsound rules in its \eqsat
      runs, but with extra checks to detect unsound merges and recover from them.
    }.
  \item \C{C}: 15 handwritten rules about the complex exponential, including rules like
  $\cis(a + b) \leftrightsquigarrow \cis(a) \cdot \cis(b)$,
  $\cis(0) \leftrightsquigarrow 1$, and $i \cdot i \leftrightsquigarrow -1$.
  \item \C{E}: 18 \textit{exploratory} rules (\autoref{sec:lift}), which
    give the definitions of $\sin$, $\cos$, and $\tan$ in terms of $\cis$, $i$, and $\pi$.
\end{itemize}

Similarly, the logarithmic, power, square root,
  and cube root functions are all defined in terms of $e^x$,
  the real exponential function:
  $e^{\log(x)} \leftrightsquigarrow x$,
  $a^b \leftrightsquigarrow e^{b \cdot \log(a)}$,
  $\sqrt{a} \leftrightsquigarrow e^{1/2 \cdot \log(a)}$, and
  $\sqrt[3]{a} \leftrightsquigarrow e^{1/3 \cdot \log(a)}$.
In the exponential domain, we have two trusted rulesets:
\begin{itemize}
  \item \C{RAT}: 141 automatically generated arithmetic rules
  \item \C{EXP}: 13 handwritten rules involving the real exponential function,
    which include rules that define
    $\log$, powers, and roots in terms of $e^x$,
  to bootstrap fast-forwarding.
\end{itemize}
For both the exponential and trigonometric domains,
  the set of known arithmetic rules can be produced by
  \enumo~via methods described in previous sections.

\paragraph*{\herbie.}

\herbie~\citep{herbie} is a widely used, open-source tool
  for improving the accuracy of floating-point expressions.
Given a mathematical expression over real numbers,
  it synthesizes a more accurate
  floating-point implementation
  using a variety of techniques
  including equality saturation.
\herbie's equality saturation-based optimization pass
  uses a set of 358 expert-written rewrite rules
  to explore many programs
  that are equivalent over the reals,
  keeping only those that have
  lower floating-point error.
\herbie's rewrite rules include many algebraic identities
  about rational arithmetic, trigonometry, and exponents.

\paragraph*{Results.}
\label{para:herbie}
\begin{figure}
  \begin{subfigure}[t]{0.49\linewidth}
    \includegraphics[width=\linewidth]{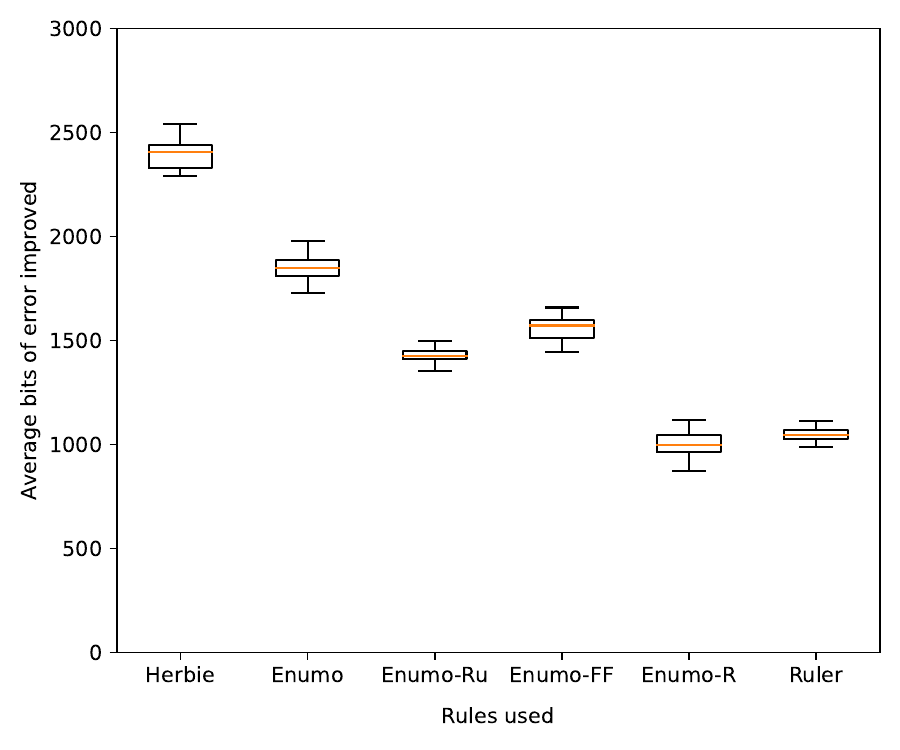}
  \end{subfigure}
  \hfill
  \begin{subfigure}[t]{0.49\linewidth}
    \includegraphics[width=\linewidth]{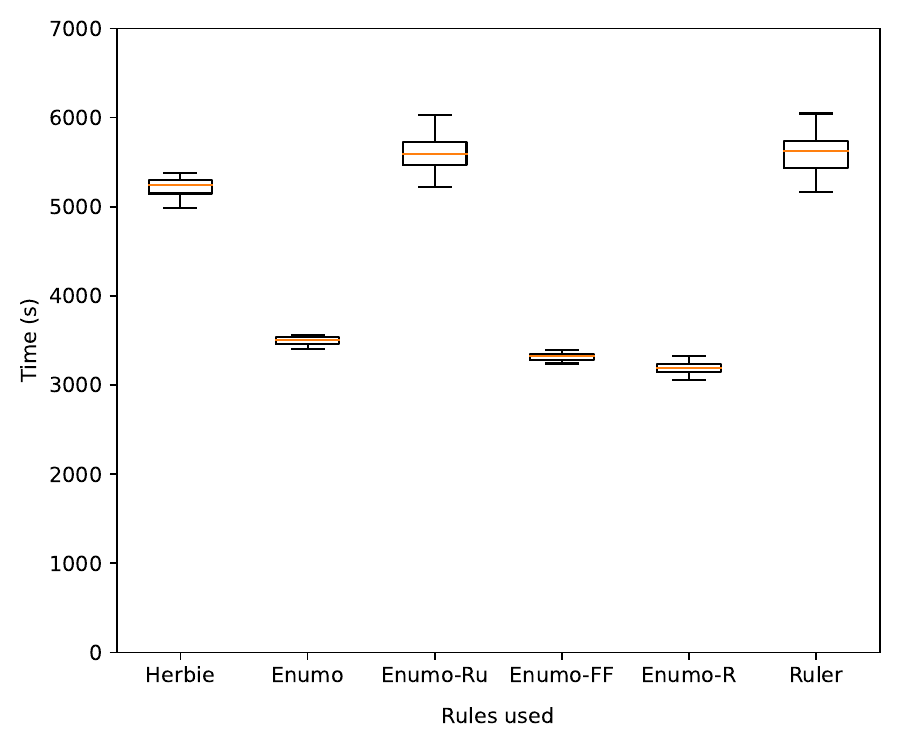}
  \end{subfigure}
  \caption{
    Comparison of different rules on \herbie's
      end-to-end performance for six different configurations:
      \herbie's default rules (\texttt{Herbie}),
      \enumo's rules (\texttt{\enumo}),
      \enumo's rules with its rational rules replaced by \ruler's rational rules (\texttt{\enumo-Ru}),
      \enumo's rules without fast-forwarded rules (\texttt{\enumo-FF}),
      \enumo's rational rules (\texttt{\enumo-R}),
      and \ruler's rules (\texttt{\ruler}).
    The two plots show
      (left) \herbie's metric for measuring accuracy (higher is better); and
      (right) \herbie's running time (lower is better).
    Each boxplot represents the results from 30 seeds,
     where each data point is obtained by summing the values
     (average error, time) over the 170 benchmarks that
     finish within the time limit.
    \enumo's rules allow \herbie to improve error
      significantly more than \ruler's rules.
  }%
  \label{fig:herbie-res}
\end{figure}

First, we wrote \enumo programs to synthesize boolean,
  rational, trigonometric (fast-forwarded),
  and exponential (fast-forwarded) rules for \herbie.
The summary of these results is in \autoref{table:herbie}.
Note that \enumo's rules derive only about a third of
  \herbie's trigonometric ruleset:
  \herbie's handwritten rules include forms beyond
  the scope of our workloads and, as discussed below,
  some unsound rules that no sound ruleset can derive.
Then, based on suggestions from \herbie's developers,
  we filtered the \herbie benchmark suite to 176
  representative benchmarks, taken from a variety of domains,
  including graphics, mathematics, and numerical analysis.
In addition, we disabled polynomial approximation to
  isolate the effects of \eqsat within \herbie.
We ran \herbie on the benchmarks under six different configurations:
\begin{itemize}
  \item \foottt{\herbie}: \herbie's default configuration.
  \item \foottt{\enumo}: \enumo's rules, including boolean, rational, trigonometric, and exponential.
  \item \foottt{\enumo-Ru}: \enumo's rules with its rational rules replaced by \ruler's rational rules. This includes \enumo's boolean, trigonometric, and exponential rules, and also \ruler's rational rules.
  \item \foottt{\enumo-FF}: \enumo's rules without fast-forwarded rules.
  \item \foottt{\enumo-R}: \enumo's rules, including only
  boolean and rational rules.
  \item \foottt{Ruler}: We compare against \ruler's rules~\citep{ruler}
    for the rational and boolean domains.
    \ruler does not support the trigonometric and exponential domains.
\end{itemize}

We used the default node limit of 8000 nodes
  in \herbie's underlying \eqsat engine,
  i.e., upon hitting the limit,
  the engine stops applying the simplification rules.
On 6 benchmarks, \herbie does not finish
  within 300 seconds; we discard these.
For all six configurations,
  we ran \herbie on 30 seeds.

\autoref{fig:herbie-res} shows the results of running
  \herbie with one boxplot for each ruleset configuration.
The left plot
  shows the improvement in accuracy achieved by \herbie,
  measured in \herbie's ``bits of error'' metric:
  the number of ``incorrect'' bits
  in the binary representation of the floating-point
  result against a high-precision oracle
  (higher improvement is better).
The right plot shows the average running time of \herbie;
  \enumo's rules consistently yield faster runs
  than \ruler's rules.
\herbie's handwritten ruleset (\herbie) finds the
  most accurate programs, followed by \enumo's rules.
This experiment offers two key takeaways:
  (1) \ff is valuable in practice, and (2) \ff and
  guided search combined lead to better rulesets than
  exhaustive synthesis alone.

Disabling trigonometric and exponential rules (\texttt{\enumo-FF}),
  \emph{but leaving the rules needed to bootstrap fast-forwarding},
  results in a significant drop in accuracy,
  demonstrating that fast-forwarding is necessary in
  the face of resource limits.
By definition, the rulesets from both \texttt{\enumo} and \texttt{\enumo-FF} have equal proving power,
  but \autoref{fig:herbie-res} demonstrates
  a significant advantage for \texttt{\enumo} over \texttt{\enumo-FF} in practice.
Using only the rational rules (\texttt{\enumo-R}) also results in lower accuracy,
  showing that trigonometric and exponential rewrite rules are important for \herbie.

To our surprise,
  Ruler's rational rules (\texttt{\ruler}) outperformed
  those from \enumo (\texttt{\enumo-R}).
However, we show that when combined with the rest of
  \enumo's ruleset (\texttt{\enumo}), \enumo finds more accurate programs
  faster than \ruler.
Replacing \enumo's rational rules with \ruler's (\texttt{\enumo-Ru})
  results in a significantly worse ruleset.
The reason is that the combination of \enumo's rational ruleset
  and \enumo's trigonometric and exponential rules allows \herbie
  to fix a wider range of floating-point errors.
We suspect that \enumo's rational rules (\texttt{\enumo-R}) explore a larger
  space than \ruler's (causing \herbie to exceed resource limits faster)
  without any benefit, leading to a slight accuracy loss compared to
  \ruler's rational rules.

An example of where rules from \enumo excel over rules from \ruler
  is \herbie's ``2cos''%
  \footnote{\url{https://github.com/herbie-fp/herbie/blob/d35c6a3cc7ab/bench/hamming/trigonometry.fpcore}}
  benchmark, $\cos(x + \varepsilon) - \cos x$,
  which suffers from error when $\varepsilon$ is relatively small.
With \enumo's ruleset,
  \herbie~uses the essential rewrite
  \C{(cos (+ b a)) \rewritesto ~} \\
  \C{(- (* (cos b) (cos a)) (* (sin b) (sin a)))}
  to decompose $\cos(x + \varepsilon)$, and then
  eliminates cancellation using associative rules.
The full rewrite is
  $\cos(x + \varepsilon) - \cos x
  \rightsquigarrow
  \cos x \cdot (\cos\varepsilon + -1) + \sin x \cdot (-\sin\varepsilon)$.

However, \herbie still finds more accurate programs
  with its handwritten ruleset.
One significant reason for this difference is that
  \herbie often relies on unsound
  division, trigonometry, and exponentiation rules in order
  to eliminate sources of errors
  such as cancellation without checking if
  such transformations are correct for all arguments.
For example,
  for the benchmark ``2sin''%
  \footnote{\url{https://github.com/herbie-fp/herbie/blob/d35c6a3cc7ab/bench/hamming/trigonometry.fpcore}},
 \herbie rewrites
 $\sin \left(x + \varepsilon\right) - \sin x$ to
 $\cos x \cdot \sin \varepsilon + ({\sin \varepsilon}^{2} \cdot \sin x)/(-1 - \cos \varepsilon)$
 using a series of rewrites including the unsound
 factoring rule
 \C{(+ a b)} \rewritesto ~ \C{(/ (- (* a a) (* b b)) (- a b))},
 which is invalid when \C{a} equals \C{b}.
In contrast, \enumo generates a conditional
  factoring rule with the side condition $(- a b) \neq 0$.
Unfortunately, \herbie is not designed to leverage conditional rules.
With \enumo, however, we can instead reify the guard syntactically
  within the rewrite itself.
\herbie can directly apply such rules, e.g.,
  \C{(+ a b)}
  \rewritesto ~
  \C{(if (- a b)
         (/ (- (* a a) (* b b)) (- a b))
         (+ a b))},
relying on other rules to simplify the condition syntactically.
\herbie's use of unsound rules and its lack of support
  for conditional rules pose a significant challenge in
  closing the gap between its handwritten rules
  and \enumo's generated rules.

\newcommand{\baselineCosImpls}{16~}
\newcommand{\baselineCosUnique}{4~}
\newcommand{\baselineCosUniqueIds}{2~}

\newcommand{\enumoCosImpls}{8~}
\newcommand{\enumoCosUnique}{5~}
\newcommand{\enumoCosUniqueIds}{3~}

\paragraph*{Megalibm.}
Next, we show how the
  ruleset inferred using \enumo for \herbie
  is also useful for
  Megalibm~\citep{megalibm}, another \eqsat
  tool that relies on numeric rewrites.
Given a transcendental operator,
  e.g., $\cos$,
  Megalibm synthesizes a set of low-level implementations
  that make different speed vs.\ accuracy tradeoffs.
A core phase in Megalibm
  is using \eqsat to discover various
  identities over such operators.

Using sound rules from
  rational, trig, and exponential \enumo programs (\autoref{table:herbie}),
  we ran the Megalibm benchmarks for $\sin$, $\cos$, and $\tan$.
Compared to Megalibm's manually developed ruleset,
  \enumo's rules would ideally discover
  at least as many \textit{unique} identities, in turn
  hopefully yielding as many or more \textit{unique}
  implementations across the speed vs.\ accuracy tradeoff space.

We detail the results for $\cos$ where
  the Megalibm baseline found the following identities:
  $\cos(x) = \cos(x + 4\pi)$,
  $\cos(x) = \cos(x + \pi + \pi)$,
  $\cos(x) = 2\cdot2\cdot \cos(-x) ~/~ 4$, and
  $\cos(x) = (\pi+\pi) - ((\pi+\pi) - \cos(-x))$.
The third and fourth identities are equivalent:
  both can be simplified to $\cos(x) = \cos(-x)$.
Similarly, the first identity
  is just two applications of the second.
Thus, the baseline only yielded 2 unique identities,
  from which Megalibm generated
  \baselineCosUnique implementations
  with differing speed and accuracy.
With \enumo-generated rewrite rules,
  Megalibm produces the following identities:
  $\cos(x) = \cos(x + (\pi + \pi))$,
  $\cos(x) = \cos(-x) \cdot 2 / 2$,
  $\cos(x) = \cos(\pi - x) - (\cos(\pi - x) \cdot 2)$,
  $\cos(x) = \cos(\pi - x) \cdot (- (\cos(\pi - x))^0)$, and
  $\cos(x) = -\cos(\pi - x)$.
Here, the third, fourth, and fifth identities are equivalent,
  yielding \enumoCosUniqueIds unique identities,
  which Megalibm used to find \enumoCosUnique unique implementations.
For $\tan$, Megalibm was able to use
  \enumo-generated rules to find
  \textit{new} identities of $\tan$
  where the baseline ruleset did not derive any.

\begin{figure}[t!]
  \centering
  \begin{subfigure}[T]{0.41\linewidth}
  \includegraphics[width=\linewidth]{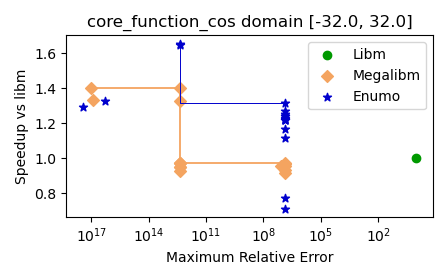}
  \label{fig:megalibm-impls}
\end{subfigure}
  \hfill
  \begin{subfigure}[T]{0.58\linewidth}
    \centering
    \vspace{2em}
    \footnotesize
  \begin{tabular}{ccccc}
    \multirow{2}{*}{Fn} & \multicolumn{2}{c}{Megalibm} & \multicolumn{2}{c}{\enumo} \\ \cline{2-5}
    & \multicolumn{1}{c}{Unique Impls} & Unique Ids & \multicolumn{1}{c}{Unique Impls} & Unique Ids \\ \hline
    sin & \multicolumn{1}{c}{7} & 2 & \multicolumn{1}{c}{5} & 3 \\
    cos & \multicolumn{1}{c}{4} & 2 & \multicolumn{1}{c}{5} & 3 \\
    tan & \multicolumn{1}{c}{\color{red}{0}} & \color{red}{0} & \multicolumn{1}{c}{3} & 1  \\
    \end{tabular}
\end{subfigure}
  \caption{Megalibm analysis.
  (Left)
  The Pareto curve shows
    the implementations Megalibm found for
    cosine over the interval [-32.0, 32.0], with
  results normalized to the GNU libm implementations.
  Points up and to the right are better (faster and less error).
  Uniqueness is judged by clusters of performance.
  (Right)
  The number of unique identities and
    implementations generated with Megalibm's original rules
    and \enumo-synthesized rules.
  Notably,
    Megalibm found no identities or implementations for
    $\tan$, but \enumo did.
  }%
  \label{fig:megalibm-results}
\end{figure}

\autoref{fig:megalibm-results} shows
  Megalibm's estimates of the speed vs.\ accuracy
  tradeoffs for implementations of $\cos$ generated by
  the \enumo-generated and manually developed baseline rulesets.
The table summarizes how,
  across $\sin$, $\cos$, and $\tan$,
  rules from \enumo always produced
  more unique identities,
  which typically also led to more unique implementations---except
  $\sin$, where
  the baseline yielded two extra implementations.
\enumo's rulesets
  can be applied across different tools in
  related domains and perform similarly to or better
  than manually developed expert rulesets.

\subsubsection{Geometric domain}
\label{subsec:geo-domain}
\newcommand{\Caddy}{\textit{Caddy}\xspace}
\newcommand{\CoreCaddy}{\textit{Core Caddy}\xspace}

\sz~\citep{szalinski} is an \eqsat-driven
  tool that shrinks 3D CAD (Computer-Aided Design) programs
  by performing rewrites
  over a language called \Caddy.
  \Caddy expresses CAD programs with
  primitives for
  constructive solid geometry (e.g., \C{Cube}, \C{Scale}, \C{Union}),
  basic rational arithmetic,
  various list constructors,
  and inverse transformations.
\sz shrinks \Caddy programs using
  two rulesets:
  a set of CAD
  identities
  and a set of custom procedural rewrite rules
  which discover opportunities to use inverse
  transformations.

We focus on synthesizing
  the CAD identities,
  which help \sz expose
  hidden structure in its input programs.
Synthesizing these identities using
  traditional rule inference algorithms
  requires a full CAD interpreter,
  which is difficult to implement and
  too slow for use in rule synthesis.
Instead, we leverage \enumo's \ff approach.

\paragraph*{Synthesizing CAD identities.}
Solid geometry can be
  represented mathematically via
  a function representation (F-Rep).
An F-Rep is a function $f(x, y, z)$
  which interprets an arithmetic expression
  over $x$, $y$, and $z$ as the geometric solid
  defined where $f$ is positive~\citep{frep}.
For example, the unit sphere can be represented
  by $1 - x^2 - y^2 - z^2$.

We use \enumo to synthesize a set of CAD
  identities by
  fast-forwarding from a small set of 5
  translational rules from CAD to F-Rep
  combined with 15 rules over the F-Rep domain.
Our prior rules for F-Rep included 6 rules
  over rationals and 9 substitution rules
  over operators that allowed it to express \Caddy
  transformations.
Here are two examples of these rules:
\[
  \mathit{Sphere}(r) \rewritesto  1 -  \left(\frac{x}{r}\right)^2 - \left(\frac{y}{r}\right)^2 - \left(\frac{z}{r}\right)^2
  \qquad
  \mathit{Scale}([w, h, l], e) \rewritesto e\left[ x \mapsto \frac{x}{w}\right]\left[ y \mapsto \frac{y}{h}\right]\left[ z \mapsto \frac{z}{l}\right]
\]
Using \enumo's guided search and fast-forwarding,
  we synthesized a set of 10 large bidirectional
  CAD identities of up to 13 atoms:
\begin{enumerate}
  \item \foottt{(Trans (Vec3 0 0 0) a) \rewritesboth \xspace a}
  \item \foottt{(Scale (Vec3 1 1 1) a) \rewritesboth \xspace a}
  \item \foottt{(Cylinder (Vec3 b b b) a true) \rewritesboth \linebreak
    (Scale (Vec3 b b b) (Cylinder (Vec3 1 1 1) a true))}
  \item \foottt{(Sphere b a) \rewritesboth \xspace (Scale (Vec3 b b b) (Sphere 1 a))}
  \item \foottt{(Scale (Vec3 f e d) (Cube (Vec3 c b a) false)) \rewritesboth \linebreak
    (Scale (Vec3 (* f c) (* e b) (* d a)) (Cube (Vec3 1 1 1) false))}
  \item \foottt{(Cube (Vec3 (* f e) (* d c) (* b a)) false) \rewritesboth \linebreak
    (Scale (Vec3 f d b) (Cube (Vec3 e c a) false))}
  \item \foottt{(Trans (Vec3 g f e) (Scale (Vec3 d c b) a)) \rewritesboth \linebreak
    (Scale (Vec3 d c b) (Trans (Vec3 (/ g d) (/ f c) (/ e b)) a))}
  \item \foottt{(Scale (Vec3 g f e) (Trans (Vec3 d c b) a)) \rewritesboth \linebreak
    (Trans (Vec3 (* d g) (* c f) (* b e)) (Scale (Vec3 g f e) a))}
  \item \foottt{(Scale (Vec3 g f e) (Scale (Vec3 d c b) a)) \rewritesboth \linebreak
    (Scale (Vec3 (* g d) (* f c) (* e b)) a)}
  \item \foottt{(Trans (Vec3 (+ g f) (+ e d) (+ c b)) a) \rewritesboth \linebreak
    (Trans (Vec3 g e c) (Trans (Vec3 f d b) a))}
\end{enumerate}

The workload we used is shown in \autoref{fig:szalinski-workload}.

\paragraph*{Results.}
We evaluate \sz's performance
  on a set of
  benchmarks taken from the paper (Table 2 of \citealt{szalinski})\footnote{
    \citet{szalinski} states that
     these benchmarks were decompiled using the
     Reincarnate~\citep{reincarnate} mesh decompiler.}.
The results are shown in \autoref{fig:szalinski-results}.
Using \enumo's synthesized rules for CAD identities, \sz is able
  to shrink input benchmarks by 87\% on average, while the original, handwritten
  rules shrink input benchmarks by 90\% on average.
Upon closer inspection, we found that a subset of the \enumo-synthesized
  rules matches the performance of \sz's handwritten CAD identities.
There are some rules in the \enumo ruleset that
  cannot be derived from \sz's CAD identities.
However, we posit that these additional rules
  are not very useful for the \sz benchmark tests, so their
  presence in the \enumo ruleset leads to worse
  performance at low node limits.

\begin{figure}

  \begin{lstlisting}[
    name=szalinski-workload-code,
    language=Python,
    basicstyle=\footnotesize\ttfamily,
    numbers=left,
    xleftmargin=2.5em]
  def iter_szalinski(n):
    lang = { (AFFINE VEC SOLID) (Cube VEC) (Cylinder VEC) (Sphere SCALAR) }
    return iter_metric(lang, "SOLID", Depth, n)
             .plug("VEC", { (Vec3 0 0 0) (Vec3 1 1 1) (Vec3 a a a) (Vec3 a b c)
                           (Vec3 d e f) (Vec3 (BOP a d) (BOP b e) (BOP c f)) })
             .plug("AFFINE", { Scale Trans })
             .plug("SCALAR", { a 1 })
             .plug("BOP", { + * / })

  cad_idents = []
  for i in [2, 3]:
      wkld = iter_szalinski(i)
      chosen = fast_forward(wkld, frep_rules, cad_idents)
      cad_idents.extend(chosen)
  \end{lstlisting}
  \caption{
    An \enumo program for learning CAD identities. \texttt{iter_szalinski}
    is a function that constructs workloads for \Caddy expressions.
  }
  \label{fig:szalinski-workload}
  \label{fig:szalinski-recipe}
\end{figure}

\begin{table}[h]
    \footnotesize
  \begin{tabular}{lccc}
    Program Id & No Identities & \sz's Identities & \enumo's Identities \\
\midrule
TackleBox & 79\% (60) & 91\% (26) & 85\% (41) \\
SDCardRack & 72\% (57) & 87\% (26) & 86\% (28) \\
SingleRowHolder & 84\% (31) & 92\% (16) & \textbf{92\% (16)} \\
CircleCell & 61\% (31) & 80\% (16) & \textbf{80\% (16)} \\
CNCBitCase & 88\% (27) & 93\% (15) & 93\% (16) \\
CassetteStorage & 81\% (27) & 89\% (15) & \textbf{89\% (15)} \\
RaspberryPiCover & 90\% (27) & 96\% (12) & \textbf{96\% (12)} \\
ChargingStation & 81\% (27) & 89\% (15) & 82\% (25) \\
CardFramer & 52\% (83) & 76\% (42) & \textbf{76\% (42)} \\
HexWrenchHolder & 90\% (31) & 95\% (16) & 84\% (52) \\
\midrule
Average & 80\% (40.1) & 90\% (19.9) & 87\% (26.3) \\
    \end{tabular}

\caption{End-to-end evaluation of
  \sz{} (Table 2 of \citealt{szalinski})
  using CAD identities from \enumo.
The inputs are flat CAD programs from  Reincarnate~\citep{reincarnate}.
For each set of identities,
  we show the percentages by which the initial program's AST sizes decreased
  (higher is better), and in parentheses, the output AST sizes (lower is better).
\texttt{No Identities}, \texttt{\sz's Identities}, and \texttt{\enumo's Identities} show the results of running \sz
  without CAD identities, with all identities enabled, and with
  \enumo's synthesized CAD identities in place of the original ones, respectively.
  \enumo's CAD identities exactly matched the
  performance of Szalinski's on 5 of 10 inputs.}

\label{fig:szalinski-results}
\end{table}

\autoref{fig:szalinski-results} shows that
  \enumo's CAD identities are able to closely match the performance
  of the handwritten CAD identities.
Fast-forwarding is effective at synthesizing rules for
  constructive geometry,
  which is a domain containing rewrite rules with term sizes that
  cannot be exhaustively enumerated,
  for which writing an interpreter is challenging,
  and which prior work did not support.

\subsection{Extensibility offered by \enumo}
\label{sec:case-study-llm}
We have demonstrated that \enumo's core operators enable more
  flexible and incremental rule inference strategies than prior work.
In this section, we show how to customize and extend the \enumo DSL to
  support entirely new techniques for rule synthesis.
Given the increasing popularity of large language models (LLMs) for synthesis
  tasks~\citep{hysynth,asap,shraddha,paper1,paper2,paper3,paper4},
  we extend \enumo to support rule inference using LLMs.
For \eqsat applications, rule soundness is paramount: even a single
  unsound rule can propagate merges through the \egraph.
As such, LLM-generated rulesets must be treated as untrusted and
  should not be used directly without processing to eliminate unsound rules.
Since \enumo already includes operators for processing untrusted rule candidates
  (\C{minimize} and \C{is_valid}), we find it to be a natural fit for
  working with untrusted LLM-generated rulesets.
Concretely, we added two new operators to \enumo{}: \C{workload_from_llm} and
  \C{ruleset_from_llm}.
The input to each operator is a natural-language prompt string\footnote{
  The prompts sent to the models are under \C{jfp/} at tag
  \C{jfp2026} of the repository
  (\url{https://github.com/uwplse/ruler/tree/jfp2026}).
}
  and the output is a \C{Workload} or \C{Ruleset}, respectively.
We measure how easily LLMs can be incorporated into \enumo and how the
  resulting operators compose with the existing ones.
Comparing the results from different LLMs is beyond the scope of this work,
  as is comparing LLM performance with different prompts for the same task.
To reduce variance of LLM performance, we use several models\footnote{
  Gemini 3.6 Flash, GPT 5.6 Luna, and Claude Sonnet 5}, querying
  each model twice, and aggregating all of the responses together.
These two extensions to \enumo required only 249 lines of Rust
  code to implement\footnote{The LLM client module plus the two
  operators' constructors and the workload's term filter, at the
  tag cited above.}, including only minimal changes to existing
  code.

\subsubsection{Ruleset synthesis}
In the first case study, we use LLMs for end-to-end ruleset synthesis
  in three domains: Halide, Trigonometric, and Exponential.
In each case, our prompt
  explained the task of rule inference at a high level
  and specified the valid values and operators in the domain
  (the domain descriptions in \autoref{table:llmrulesets}).
The prompt also included example rules and listed classes of rules to cover:
  for example, commutativity, associativity, and distributivity for Halide;
  angle-sum and double-angle identities for Trigonometric;
  and product, quotient, and power laws for Exponential.
Finally, it stressed that every rule must be sound wherever both sides
  are defined and asked for plain text, one rule per line, with no commentary.
We parse rules out of the LLM response, discarding any malformed
  rules, and treat the resulting ruleset as \textit{untrusted rule candidates},
  which we verify as explained below.
Since we aggregate responses from several models and queries,
  there are many redundant rules generated, which
  we eliminate using \enumo's \C{minimize} operator.

\paragraph*{Verifying LLM-generated rule candidates.}
For Halide, we validate the LLM-generated rule candidates using
  \enumo's \C{is_valid} operator,
  which checks the rule by encoding it as an SMT query.
Empirically, we find that unsound rules from LLMs are a genuine problem:
  in the Halide domain, 36 of the 977 LLM-generated rule candidates were
  found to be unsound.
While this is a relatively small percentage of the rules, in \eqsat settings,
  the presence of a single unsound rule can invalidate all of the
  equalities represented in the \egraph.
For Trigonometric and Exponential, validity
  is undecidable in general~\citep{dreal},
  so SMT-based validation does not apply.
Instead, we reuse an existing \enumo operator, \C{can_derive},
  to check soundness by derivability,
  following the same intuition as our \ff algorithm (\autoref{sec:lift}).
Specifically, we attempt to derive each
  LLM-synthesized rule candidate using
  the known rules for each domain: \C{R}\footnote{
    While unsound rules were acceptable in \autoref{sec:lift} where we could ensure
    that no undefined terms were represented in the \egraph, in this setting,
    we exclude the 13 arithmetic rules involving division that are unsound at zero.
  } \C{+ C + E} for Trigonometric and \C{RAT + EXP}
  for Exponential (\autoref{sec:lift}).
Assuming soundness of these trusted rules,
  if a rule candidate can be derived
  from them,
  then the rule candidate is sound.
If a rule candidate cannot be derived, it is filtered out,
  even though it may be sound.

\paragraph*{Eliminating redundant LLM-generated rules.}
Once we have filtered out potentially unsound rule candidates,
  we address the problem
  of redundant rules.
We use \enumo's \C{minimize} operator (\autoref{subsec:ruleset-sem}),
  which reduces a ruleset with respect to a set of prior rules.
For Halide, we consider three sets of prior rules: \C{None} (the empty ruleset),
  \C{A5} (rules synthesized by exhaustively enumerating terms up to size 5),
  and \C{ENUMO} (rules synthesized using a custom \enumo
  program, \autoref{subsec:halideeval}).
For Trigonometric, we use the arithmetic rules \C{R} without the
  division rules (as in \autoref{table:herbie}) as prior rules for
  minimization; for Exponential, the rational rules \C{RAT} alone.

\begin{table}[t]
\resizebox{\textwidth}{!}{%
\begin{tabular}{|l|l|l|l|c|l|}
\hline
Ruleset Name & Domain                         & Domain Description                                                                                                                                                                                 & Prompt Rules & \multicolumn{1}{l|}{Validation}                                                     & Prior Rules         \\ \hline
LLM-1        & \multirow{6}{*}{Halide}        & \multirow{6}{*}{\begin{tabular}[c]{@{}l@{}}Values: integers\\ Unary Operators: -, !\\ Binary Operators: \C{<}, \C{<=}, ==, !=, \&\&, \C{||}, \^{}, +, -, *, min, max\\ Ternary Operators: select\end{tabular}} & -            & \multirow{6}{*}{SMT}                                                                & -                   \\
LLM-2        &                                &                                                                                                                                                                                               & LLM-1        &                                                                                     & LLM-1               \\
LLM-A5-1     &                                &                                                                                                                                                                                               & -            &                                                                                     & A5                  \\
LLM-A5-2     &                                &                                                                                                                                                                                               & LLM-A5-1     &                                                                                     & A5 + LLM-A5-1       \\
LLM-ENUMO-1  &                                &                                                                                                                                                                                               & -            &                                                                                     & ENUMO               \\
LLM-ENUMO-2  &                                &                                                                                                                                                                                               & LLM-ENUMO-1  &                                                                                     & ENUMO + LLM-ENUMO-1 \\ \hline
LLM-1        & \multirow{3}{*}{Trigonometric} & \multirow{3}{*}{\begin{tabular}[c]{@{}l@{}}Values: real numbers, PI\\ Unary operators: -, sin, cos, tan, sqr\\ Binary operators: +, -, *, /\\\end{tabular}}                                   & -            & \multirow{3}{*}{\begin{tabular}[c]{@{}c@{}}Derivability from\\ R + C + E\end{tabular}} & R                  \\
LLM-2        &                                &                                                                                                                                                                                               & LLM-1        &                                                                                     & R + LLM-1          \\
             &                                &                                                                                                                                                                                               &              &                                                                                     &                     \\ \hline
LLM-1        & \multirow{3}{*}{Exponential}   & \multirow{3}{*}{\begin{tabular}[c]{@{}l@{}}Values: real numbers\\ Unary operators: -, exp, log, sqrt, cbrt\\ Binary operators: +, -, *, /, pow\\\end{tabular}}                                  & -            & \multirow{3}{*}{\begin{tabular}[c]{@{}c@{}}Derivability from\\ RAT + EXP\end{tabular}} & RAT                  \\
LLM-2        &                                &                                                                                                                                                                                               & LLM-1        &                                                                                     & RAT + LLM-1          \\
             &                                &                                                                                                                                                                                               &              &                                                                                     &                     \\ \hline
\end{tabular}%
}
\caption{Description of how each ruleset is generated using LLMs.}
\label{table:llmrulesets}
\end{table}

In this experiment, we use LLMs
  in an iterative manner to
  incrementally construct rulesets.
After generating rules, we prompt
  the LLMs again in order to find missing rules.
The second prompt restates the domain, lists the rules already in the
  ruleset (the prompt rules in \autoref{table:llmrulesets}),
  and asks for sound rules missing from that list, excluding trivial
  variants such as renamed variables or swapped arguments of
  commutative operators.
This results in a new set of rule candidates,
  which we validate and minimize again
  for each domain.
For this second round of minimization,
  the rules generated in the first round
  are also included as prior rules.
\autoref{table:llmrulesets} summarizes the rulesets generated for this case study.

For each ruleset, we measured the
  \lhsandrhs derivability compared to two baselines: \C{ENUMO},
  the rules synthesized using \enumo programs as
  outlined in \autoref{subsec:halideeval} and \autoref{subsec:eval-ff},
  and \C{HALIDE} (for Halide) or \C{HERBIE} (for Exponential and Trigonometric).
As in \autoref{table:herbie}, every deriving ruleset in the Trigonometric
  and Exponential matrices is extended with \C{R} or \C{RAT}, respectively;
  the LLM rulesets were minimized against these rules and are not
  meaningful without them.
For the Halide domain, we also compare to \C{A5} as a third baseline.
The full results are shown in \autoref{table:llm1}.

As expected, we find that LLMs are powerful tools for ruleset synthesis.
Measured against the expert-written \C{HALIDE} and \C{HERBIE} baselines,
  the best LLM-synthesized ruleset outperforms guided search in
  the Halide and Exponential domains.
However, in the Trigonometric domain, the LLM-synthesized ruleset underperforms
  compared to \C{ENUMO}.
In all domains, neither technique subsumes the other:
  \C{ENUMO} rulesets contain
  rules not derivable from the LLM rulesets, and vice versa,
  indicating that they find substantively different rules.

We also find that LLMs compose well with traditional rule inference techniques.
For example, \C{LLM-A5-1} outperforms both \C{A5} and \C{LLM-1}
  against the \C{ENUMO} baseline, suggesting that \textit{composing
  traditional rule inference methods with LLMs can lead to
  better rulesets than using either technique alone}.
Further, re-prompting shows that the new operators compose with
  themselves and with the rest of the pipeline: the minimized first-round
  ruleset is embedded in the second prompt, and the second round's
  candidates are validated and minimized against the first round's rules.

\begin{table}[ht]
    \centering

    \resizebox{\textwidth}{!}{%
    \begin{tabular}{|lr|lllllllll}
    \hline
    Ruleset      & Synthesis Time (s)  & HALIDE       & A5           & ENUMO                             & LLM-1        & LLM-2       & LLM-A5-1    & LLM-A5-2    & LLM-ENUMO-1  & \multicolumn{1}{l|}{LLM-ENUMO-2} \\ \hline
    A5           & 15.7\phantom{$^{+}$} & 45.1\% (10.6) & -             & 69.3\% (26.5)                     & 53.0\% (8.1)  & 53.8\% (8.0) & 4.4\% (4.5)  & 16.5\% (5.5) & 14.9\% (3.5) & \multicolumn{1}{l|}{23.8\% (4.4)} \\
    ENUMO        & 49.2\phantom{$^{+}$} & 90.9\% (22.1) & 88.3\% (56.6) & -                                  & 74.5\% (32.6) & 74.9\% (34.6) & 44.4\% (15.5) & 48.5\% (18.7) & 17.9\% (21.2) & \multicolumn{1}{l|}{31.2\% (18.7)} \\ \cline{6-11}
    LLM-1        & 1043.7\phantom{$^{+}$} & 93.5\% (5.6)  & 64.0\% (19.8) & \multicolumn{1}{l|}{65.0\% (38.7)} &               &              &              &              &               &                                  \\
    LLM-2        & 326.8$^{+}$         & 94.3\% (4.7)  & 67.7\% (19.9) & \multicolumn{1}{l|}{68.7\% (39.5)} &               &              &              &              &               &                                  \\
    LLM-A5-1     & 1043.5\phantom{$^{+}$} & 90.1\% (38.1) & -             & \multicolumn{1}{l|}{81.2\% (25.3)} &               &              &              &              &               &                                  \\
    LLM-A5-2     & 586.2$^{+}$         & 89.4\% (36.5) & -             & \multicolumn{1}{l|}{81.3\% (17.2)} &               &              &              &              &               &                                  \\
    LLM-ENUMO-1  & 1043.5\phantom{$^{+}$} & 91.2\% (24.6) & 91.5\% (11.8) & \multicolumn{1}{l|}{-}             &               &              &              &              &               &                                  \\
    LLM-ENUMO-2  & 734.7$^{+}$         & 94.2\% (22.9) & 90.8\% (12.9) & \multicolumn{1}{l|}{-}             &               &              &              &              &               &                                  \\ \cline{1-5}
    \end{tabular}%
    }

    \vspace{1em}

    \begin{minipage}{0.45\textwidth}
        \centering
        \resizebox{\textwidth}{!}{%
        \begin{tabular}{|lr|llll}
        \hline
        Ruleset     & Synthesis Time (s)  & HERBIE       & ENUMO                             & LLM-1        & \multicolumn{1}{l|}{LLM-2}        \\ \hline
        ENUMO       & 4.6\phantom{$^{+}$} & 43.9\% (2.7) & -                                 & 89.3\% (1.0) & \multicolumn{1}{l|}{90.6\% (1.0)} \\ \cline{5-6}
        LLM-1       & 516.0\phantom{$^{+}$} & 52.4\% (1.7) & \multicolumn{1}{l|}{65.0\% (0.5)} &              &                                   \\
        LLM-2       & 399.3$^{+}$         & 53.7\% (1.5) & \multicolumn{1}{l|}{70.0\% (0.5)}  &              &                                   \\ \cline{1-4}
        \end{tabular}%
        }
    \end{minipage}
    \hfill
    \begin{minipage}{0.45\textwidth}
        \centering
        \resizebox{\textwidth}{!}{%
        \begin{tabular}{|lr|llll}
        \hline
        Ruleset  & Synthesis Time (s)  & HERBIE        & ENUMO                             & LLM-1        & \multicolumn{1}{l|}{LLM-2}        \\ \hline
        ENUMO    & 692.4\phantom{$^{+}$} & 35.6\% (14.0) & -                                 & 72.7\% (1.4) & \multicolumn{1}{l|}{75.0\% (1.4)} \\ \cline{5-6}
        LLM-1    & 464.3\phantom{$^{+}$} & 13.3\% (1.5)  & \multicolumn{1}{l|}{44.6\% (1.1)} &              &                                   \\
        LLM-2    & 770.0$^{+}$         & 13.3\% (1.7)  & \multicolumn{1}{l|}{44.6\% (1.1)}  &              &                                   \\ \cline{1-4}
        \end{tabular}%
        }
    \end{minipage}

    \caption{Derivability matrices for our LLM rule synthesis case study.
    Halide (Top), Exponential (Left), and Trigonometric (Right).
    Derivability times in seconds are shown in parentheses.
    Reprompted rulesets show only the second round synthesis times
    (marked with $^{+}$) and do not include the first
    round's synthesis time.
    Cell values show the \lhsandrhs derivability results of using the
    row ruleset to derive the column ruleset.
    For Exponential and Trigonometric, the row ruleset is extended with
    \C{RAT} or \C{R} before deriving, as in \autoref{table:herbie}.
    A dash marks cells where the column ruleset is contained in the row ruleset,
    so derivability holds by construction.
    Derivability between LLM rulesets is not included because
    the goal is not to compare LLM-synthesized rulesets against each other.
    }
    \label{table:llm1}
\end{table}

\subsubsection{Workload synthesis}
\label{subsec:case-study-workload}
The second case study explores the use of LLMs for term enumeration.
Rather than prompting the model for complete rulesets, we prompt the model
  to enumerate terms from the domain to make an \enumo workload.
For this case study, we focus on the Halide domain.
The prompt specifies the constants, variables, and operators that terms may use
  and gives several example terms.
It also explains that
  rules will be inferred from pairs of equivalent terms in the workload,
  so the terms should come in clusters that are likely to be equivalent,
  vary in size and nesting depth, and should cover all of the operators and their combinations.
The \enumo \C{workload_from_llm} operator then constructs an \enumo \C{Workload}
  from the valid terms in the LLM response,
  skipping any malformed s-expressions
  or invalid terms.
We also filter out any terms that use variables other than those specified
  in order to keep the cvec size manageable\footnote{cvec length grows exponentially
   with the number of variables in the \C{Workload}.}.

For the Halide domain,
  our prompt tells each model to generate at least 1000 terms.
Across two queries per model, the responses contained
  3407 (Gemini 3.6 Flash), 3195 (GPT 5.6 Luna), and 2111 (Claude Sonnet 5) terms,
  which aggregated to 5952 unique terms, 5878 of them valid.
In total, generating this workload took 973.3 seconds.

To synthesize rules from the workload,
  we follow the typical \enumo pipeline:
  convert the workload to an \egraph,
  compress the \egraph using prior rules,
  find rule candidates using cvec matching, and
  minimize the rule candidates
  to eliminate redundant and unsound rules.
The LLM-synthesized workload composes seamlessly into the
rest of the rule inference pipeline, enabling entirely new
strategies for rule inference without modifying any of \enumo's other operators.

We consider four sets of prior rules:
  \C{None}, \C{A5}, \C{ENUMO} as in
  the first case study, and also
  \C{LLM-2}, which is the LLM-generated ruleset from the
  first case study for Halide.
For each ruleset, we measured the \lhsandrhs derivability compared to
  three baselines: \C{HALIDE}, \C{A5}, and \C{ENUMO}.
The full results are shown in \autoref{table:llm2}.

As in the first case study, we find that incorporating LLMs can lead to more
  powerful rulesets than using either technique alone.
Inferring rules directly from the LLM-generated workload yields 2181 rules,
  but is quite slow (roughly 1.5 hours), with the majority of the time
  spent minimizing nearly 200,000 rule candidates.
Furthermore, the resulting ruleset is quite weak, deriving only 28.1\% of
  the \C{HALIDE} baseline.
However, it is fast and easy to use \enumo to generate a set of rules
  by exhaustively enumerating terms up to size 5 (\C{A5}).
If we use these rules as a starting point for rule inference using \C{LLM-W},
  we drastically reduce the number of candidates under consideration to only 7195,
  and minimization finishes in under a minute, resulting in 369 rules.
Perhaps surprisingly, these rules are also much more powerful than
  the directly-synthesized ruleset, deriving 69.7\% of \C{HALIDE}'s.
This suggests a promising strategy for workload-driven rule inference:
  use traditional techniques for small rules, where exhaustive and guided search
  are fast and reliable, then use LLMs to sample terms that are harder to find using search.

Notably, \C{LLM-W-LLM-2}, which was entirely synthesized by LLMs,
  with no domain expert guidance,
  outperforms \C{ENUMO} against the \C{HALIDE} baseline,
  suggesting that incorporating LLMs into rule inference pipelines could
  lower the barrier to producing high-quality rulesets.

\begin{table}[h]
\footnotesize
\resizebox{\textwidth}{!}{%
\begin{tabular}{|l|llllll|}
\hline
Ruleset     & \multicolumn{1}{l}{\# Prior Rules} & \multicolumn{1}{l}{\# New Rules} & \multicolumn{1}{r}{Rule Synthesis Time (s)} & HALIDE      & A5          & ENUMO        \\ \hline
A5          & 0             & 480                         & 15.7\phantom{$^{+}$}         & 45.1\% (10.6) & -             & 69.3\% (26.5) \\
ENUMO       & 0             & 840                         & 49.2\phantom{$^{+}$}         & 90.9\% (22.1) & 88.3\% (56.6) & -              \\
LLM-2       & 0             & 370                         & 1370.4\phantom{$^{+}$}       & 94.3\% (4.7)  & 67.7\% (19.9) & 68.7\% (39.5)  \\
LLM-W       & 0             & 2181                        & 5510.8$^{+}$                 & 28.1\% (3291.6)  & 35.8\% (2359.6)  & 25.4\% (6965.6)   \\
LLM-W-A5    & 480           & 369                         & 49.1$^{+}$                   & 69.7\% (1006.6) & -             & 74.4\% (214.8)  \\
LLM-W-ENUMO & 840           & 332                         & 74.9$^{+}$                   & 91.6\% (783.1) & 90.4\% (608.7) & -              \\
LLM-W-LLM-2 & 370           & 200                         & 27.9$^{+}$                   & 93.1\% (40.9)  & 78.1\% (48.4) & 73.9\% (99.8) \\ \hline
\end{tabular}%
}
\caption{
  Derivability comparison between rulesets synthesized from an LLM-generated workload.
  Cell values show the \lhsandrhs derivability results of using the
    row ruleset to derive the column ruleset.
  Each row ruleset consists of its prior rules together with the new rules
    synthesized on top of them, and derivability is measured using both.
  Derivability times in seconds are shown in parentheses.
  Dashes indicate cells where the column ruleset is contained within the row ruleset,
    and derivability is guaranteed by construction.
  The time to synthesize \texttt{LLM-2} includes the LLM query time.
  The synthesis times for the \texttt{LLM-W} rulesets (marked with $^{+}$) do not include the time to
  generate the workload from the LLMs, which was 973.3s.}%
\label{table:llm2}
\end{table}

 These two case studies show the extensibility of
 \enumo as new rule inference techniques emerge.
Adding support for LLMs to the \enumo implementation was straightforward,
  requiring minimal changes to existing code.
The new operators compose correctly with existing \enumo operators,
  as shown in \autoref{subsec:case-study-workload}, where we use the
  new \C{workload_from_llm} operator to construct a workload, leaving
  the rest of the rule inference pipeline unchanged.
The operators can also be composed with existing operators in novel ways,
  as in the Trigonometric and Exponential case studies, where we
  use \enumo's \C{can_derive} operator to determine validity of the LLM-generated
  rule candidates.

\paragraph*{Threats to validity.}
Our claims concern the ease of extending \enumo and the composability of
  the resulting operators, not the performance of LLMs relative to guided
  search, which varied widely across domains (\autoref{table:llm1}).
In these case studies, we did not attempt to
  compare performance between LLM models,
  nor did we rigorously explore how prompt
  engineering impacts the resulting rulesets.
Both are interesting directions of future work.
Due to the nondeterministic nature of LLMs,
  the specific rules and terms generated differ from run to run.

\subsection{Ruleset manipulation with \enumo}
\label{sec:case-study-bv}

\begin{table}[h]
\footnotesize
\begin{tabular}{llll}
Domain  & Generated Rules (Time)      & Valid BV4 Rules (Time)     & Validated $\rightarrow$ Generated \\ \cline{1-4}
BV8     & 230 (32.78)                 &  230 (3.19)                & (100\%, 100\%) \\
BV16    & 236 (85.50)                 &  224 (8.36)                & (97\%, 97\%) \\
BV32    & 232 (191.74)                &  224 (10.81)               & (98\%, 99\%) \\
BV64    & 250 (43.70)                 &  220 (16.97)               & (93\%, 93\%) \\
BV128   & 190 (1784.14)               &  210 (38.68)               & (90\%, 91\%) \\
\end{tabular}%
\caption{
  Comparison of rule synthesis for different widths of bitvectors.
  Shown for each bitvector width are
    (i) the number of rules generated from an
    \enumo program (time in seconds) for that domain,
    (ii) the number of \enumo-synthesized BV4 rules
    that are valid in that domain (time in seconds), and
    (iii) the percentage of the generated rules
    that are derivable from the validated BV4 rules
    ( \lhs and \lhsandrhs derivability, in that order).
}%
\label{table:bv}
\end{table}

In this section, we show a case study in leveraging \enumo's operators for
  ruleset manipulation.
Here, we are interested in synthesizing rewrite rules over 4-bit bitvectors
  (BV4), and transforming those rules into a usable ruleset over
  larger bitvectors.
Synthesizing rules for small bitvectors is very fast because there are
  relatively few possible values in the domain.
Rules that work for small bitvectors are likely, but not guaranteed, to be
  valid for large bitvectors as well.
In this case study, we start by synthesizing BV4 rules using
  the same \enumo program as described in \autoref{subsec:eval-enumo}.
Then we use \enumo operators to translate the rules into the domain of larger
  bitvectors (BV8, BV16, BV32, BV64, and BV128).
Finally, we validate the rules in the new domain to find the subset of sound BV4
  rules that are still sound for larger bitvectors.
We compare these rules against rules that were synthesized directly in the
  larger bitvector domains using the same \enumo program as we used to
  synthesize BV4 rules.
The results are shown in \autoref{table:bv}.
Validating BV4 rules is much
  faster than synthesizing rules from scratch
  (38 seconds vs.\ 29 minutes for BV128)
  and still yields good results.
Across all bitvector sizes, the validated BV4 rules retained at least 90\% of
  the proving power of the directly generated rules, and in the case of
  8-bit bitvectors, the validated BV4 rules had equal proving power.

\section{Developer experience with \enumo}
\label{sec:devexp}

In this section, we report on the developer experience of using
  \enumo to generate rulesets for the trigonometric domain used in \autoref{subsubsec:numbers}
  and the Halide domain used in \autoref{subsec:halideeval}.

\subsection{Developing custom workloads for trigonometry}
\label{subsec:trigexperience}

We describe the process of developing, in \enumo, the workloads
  we used in \autoref{subsubsec:numbers}.
The goal is to synthesize a set of rewrite rules that will perform well in
  \herbie, a rewrite-based tool for improving the accuracy of floating-point expressions.
We take an incremental approach, composing together the results of many
  invocations of the \ff algorithm.
At each step, rules learned in prior steps are used to shrink the \egraph
  during candidate generation and rule minimization.

\subsubsection{Constants}
First we enumerate terms that apply
 trigonometric functions to a set of constants.
\autoref{subsec:ffoverview} briefly describes this workload.
We explain the process in more detail here.
Specifically, we are interested in (possibly fractional) multiples of $\pi$.
The goal is to find identities like
  \C{(cos 0) \rewritesboth\, (sin (/ PI 2))}.
We are careful to filter out terms that are undefined like
  \C{(tan (/ PI 2))}: this term is equivalent to \C{(/ 1 0)} since
  \C{(sin (/ PI 2))} is 1 and \C{(cos (/ PI 2))} is 0.
Having undefined terms in the \egraph can cause other rules
  over rationals to discover unsound equivalences\footnote{For example,
  we include the rule \C{(/ x x) \rewritesto\, 1}, which is not sound if
  \C{x} is 0, but it is very useful when \C{x} is nonzero. To avoid
  unsoundness from this rule, it is important to prevent undefined
  terms from being represented in the \egraph at any point.}.

The benefit of learning these first is that they can help with additional
  constant folding.
Since these terms have no free variables
  but represent non-trivial identities,
  it is useful to discover them first in a separate phase.

The concrete workload contains 22 terms and is described by this \enumo program:
\begin{lstlisting}[name=trig]
  app = { (OP VAL) }
  ops = { sin cos tan }
  consts = { 0 (/ PI 6) (/ PI 4) (/ PI 3) (/ PI 2) PI (* PI 2) }
  W1 = app
          .plug("OP", ops)
          .plug("VAL", consts)
          .filter(Filter.Not(Filter.Contains("(tan (/ PI 2))")))
          .union({0 1})
\end{lstlisting}

The set of rewrite rules we find using \ff from this workload is:
\begin{lstlisting}[escapechar=@]
  (cos (/ PI 4)) @\rewritesboth@ (sin (/ PI 4))
  1 @\rewritesboth@ (tan (/ PI 4))
  1 @\rewritesboth@ (sin (/ PI 2))
  1 @\rewritesboth@ (cos (* PI 2))
  1 @\rewritesboth@ (cos 0)
  (tan PI) @\rewritesboth@ (cos (/ PI 2))
  (sin PI) @\rewritesboth@ (sin (* PI 2))
  (tan PI) @\rewritesboth@ (tan (* PI 2))
  (sin 0) @\rewritesboth@ (sin PI)
  (tan 0) @\rewritesboth@ (tan PI)
  0 @\rewritesboth@ (tan PI)
  (sin PI) @\rewritesboth@ (tan PI)
\end{lstlisting}

\subsubsection{Even/odd symmetry and periodicity}
Next, we create a workload to
  enumerate trigonometric functions whose arguments are transformed by (1) negation,
  (2) translation by $\pm \pi$, or (3) doubling.
The goal is to learn axioms like
  evenness: \C{f (- x) \rewritesboth\, f (x)},
  oddness: \C{f (- x) \rewritesboth\, -f (x)},
  and anti-periodicity: \C{f (+ x  P) \rewritesboth\, -f (x)}.
Evenness and oddness are useful for techniques like range reduction
  when constructing math libraries,
  where an even (or odd) function only needs
  to be implemented for half the range and
  the input (or output) can be transformed accordingly.
Conveniently, anti-periodicity axioms lead to periodicity identities:
  \C{f (+ x (* 2  P)) \rewritesto\, f (x)}.

Learning these axioms early is ideal.
Not only are they important properties of trigonometric functions,
  they serve as simplifying rewrites in later workloads,
  eliminating unnecessary negations and translations.

The concrete workload contains 50 terms and is described by this \enumo program:
\begin{lstlisting}
  simple_terms = app
                    .plug("OP", ops)
                    .plug("VAL", {a  (- a)  (+ PI a)  (- PI a)  (+ a a)})
  neg_terms = {(- X)}.plug("X", simple_terms)
  W2 = W1.union(simple_terms).union(neg_terms)
\end{lstlisting}

The set of rewrite rules we obtain from this workload is:
\begin{lstlisting}[escapechar=@]
  (- (cos a)) @\rewritesboth@ (cos (+ PI a))
  (sin a) @\rewritesboth@ (sin (- PI a))
  (tan a) @\rewritesboth@ (tan (+ PI a))
  (- (tan a)) @\rewritesboth@ (tan (- a))
  (- (sin a)) @\rewritesboth@ (sin (- a))
  (cos (- a)) @\rewritesboth@ (cos a)
\end{lstlisting}

\subsubsection{Sum-of-squares}
With this workload, we introduce sum-of-squares
  terms which have the form: \\
   \C{(+ (* (f x) (f x)) (* (g y) (g y)))}.

A number of trigonometric identities
  contain the square of a trigonometric function.
At this point, the full set of terms is too large,
so we prune away terms of the form \C{(f (+ x x))}.
With this workload, we learn the Pythagorean identity.

The concrete workload contains 116 terms and is described by this \enumo program:

\begin{lstlisting}
  no_trig_2x_filter = Filter.Not(Filter.Or(
                            Filter.Contains("(sin (+ x x))")
                            Filter.Contains("(cos (+ x x))")
                            Filter.Contains("(tan (+ x x))")))
  squares = { (sqr X) }.plug("X", app).plug("OP", ops).plug("VAL", {a b})
  add = {(+ E E) (- E E)}
  sum_of_squares = add.plug("E", squares)
  W3 = W2.filter(no_trig_2x_filter).union(sum_of_squares)
\end{lstlisting}

The set of rewrite rules \enumo finds from this workload is:
\begin{lstlisting}[escapechar=@]
  (+ (* (sin a) (sin a)) (* (cos a) (cos a))) @\rewritesto@ 1
  (- (* (cos b) (cos b)) (* (cos a) (cos a))) @\rewritesboth@ (- (* (sin a) (sin a)) (* (sin b) (sin b)))
  (- (* (sin b) (sin b)) (* (cos a) (cos a))) @\rewritesboth@ (- (* (sin a) (sin a)) (* (cos b) (cos b)))
\end{lstlisting}

\subsubsection{Coangles}
With this workload, we try to prove coangle identities.
Similar to the (anti-)periodicity identities,
  these identities allow us to eliminate translations
  by a quarter period.
Unlike in previous workloads where we simply added terms
  to the previous workload,
  here we start with a fresh set of more targeted terms
  to learn the identities quickly.
The concrete workload contains 13 terms and is described by this \enumo program:
\begin{lstlisting}
  nums = { -2 -1 0 1 2 }
  sin_cos = { sin cos }
  base = { a  (+ (/ PI 2) a)  (- (/ PI 2) a)  (* 2 a) }
  simple = app.plug("OP", sin_cos).plug("VAL", base)
  W4 = simple.union(nums)
\end{lstlisting}
The rewrite rule \enumo finds from this workload is:
\begin{lstlisting}[escapechar=@]
  (cos (- (/ PI 2) a)) @\rewritesboth@ (sin a)
\end{lstlisting}

\subsubsection{Power reduction}
With this workload,
  we try to learn power reduction identities,
  which rewrite trigonometric functions raised to a
  power into terms with smaller powers.
For example, the following identity is a power reduction identity:
  \C{(* (sin x) (sin x)) \rewritesto\, (/ (- 1 (cos (* 2 x))) 2)}.

The concrete workload contains 57 terms and is described by this \enumo program:
\begin{lstlisting}
shift = { X (- 1 X) (+ 1 X) }
scale_down = { X (/ X 2) }
shifted_simple = shift.plug("X", simple)
trivial_squares = { (sqr X) }.plug("X", app).plug("OP", sin_cos).plug("VAL", { a })
shifted_simple_sqrs = shifted_simple.union(trivial_squares)
scaled_shifted_sqrs = scale_down.plug("X", shifted_simple_sqrs)
W5 = scaled_shifted_sqrs.union(nums)
\end{lstlisting}

The set of rewrite rules \enumo finds from this workload is:
\begin{lstlisting}[escapechar=@]
  (/ (- 1 (cos (* 2 a))) 2) @\rewritesboth@ (* (sin a) (sin a))
  (/ (+ 1 (cos (* 2 a))) 2) @\rewritesboth@ (* (cos a) (cos a))
  (- a a) @\rewritesboth@ (- (- a a))  # minimization failed to eliminate this identity
\end{lstlisting}
Note that the last rule shows a limitation of \enumo's minimization approach: this
  identity should ideally have been eliminated, but due to the incomplete nature of minimization,
  it ends up in the final ruleset.

\subsubsection{Product-to-sum}
Similar to the previous workload,
  this workload explores identities with
  products of trigonometric functions such as
  \C{(* (sin x) (sin y))}.

To avoid enumerating terms from the previous workload,
 we filter out any squared terms.
These axioms are more general than the ones from the
  previous workload but are harder to learn, so we learn them separately.
While the more general versions together with other rewrites
  can in theory be used to derive the ones in the previous section,
  due to practical limitations like resource bounds,
  real-world equality saturation applications can benefit from both sets of
  rules.

The concrete workload contains 373 terms and is described by this \enumo program:

\begin{lstlisting}
non_square_filter = Filter.Not(Filter.Or(
          Filter.Contains("(* (sin x) (sin x))")
          Filter.Contains("(* (cos x) (cos x))")))
two_var = app
            .plug("OP", sin_cos)
            .plug("VAL", { a b (+ a b) (- a b) })
sum_two_vars = { (+ X Y) (- X Y) }
                  .plug("X", two_var)
                  .plug("Y", two_var)
prod_two_vars = { (* X Y) }
                   .plug("X", two_var)
                   .plug("Y", two_var)
                   .filter(non_square_filter)
sum_and_prod = sum_two_vars.union(prod_two_vars)
scaled_sum_prod = scale_down.plug("X", sum_and_prod)
W6 = scaled_sum_prod.union(nums)
\end{lstlisting}

The set of axioms \enumo learns from this workload is:
\begin{lstlisting}[escapechar=@]
  (* (sin b) (sin a)) @\rewritesboth@ (/ (- (cos (- b a)) (cos (+ b a))) 2)
  (* (cos b) (cos a)) @\rewritesboth@ (/ (+ (cos (+ b a)) (cos (- b a))) 2)
  (+ b a) @\rewritesto@ (+ a b) # minimization failed to eliminate this identity
\end{lstlisting}

\subsubsection{Sums}
Finally, we find identities involving trigonometric
  functions with sums as their arguments, such as
  \C{(sin (+ x y))}.

These are related to the product-to-sum workload
  since products of trigonometric functions rewrite
  into trigonometric functions of sums and vice versa.
This workload enumerates many terms, so we impose a number
of filters, including removing double-angle terms,
  e.g., \C{(sin (+ x x))}, and squared terms, e.g., \C{(sin (* x x))},
  to keep the search tractable.

The concrete workload contains 287 terms and is described by this \enumo program:

\begin{lstlisting}
trig_no_sub_filter = Filter.Not(Filter.Or(
            Filter.Contains("(cos (- a b))")
            Filter.Contains("(sin (- a b))")))
two_x_filter = Filter.Not(Filter.Contains("(+ x x)"))
two_var_no_sub = two_var.filter(trig_no_sub_filter)
sum_of_prod = { (+ X Y) (- X Y) }
              .plug("X", prod_two_vars)
              .plug("Y", prod_two_vars)
              .filter(two_x_filter)
              .filter(non_square_filter)
W7 = two_var_no_sub.union(sum_of_prod).union(nums)
\end{lstlisting}

The set of axioms \enumo learns from this workload is:
\begin{lstlisting}[escapechar=@]
  (- (* (cos b) (cos a)) (* (sin b) (sin a))) @\rewritesboth@ (cos (+ b a))
  (sin (+ b a)) @\rewritesboth@ (+ (* (sin b) (cos a)) (* (sin a) (cos b)))
\end{lstlisting}

Note that the incremental approach to rule inference is crucial.
Rules learned at each step are used as prior rules in subsequent steps.
It is possible to run \ff with a single workload composed of all of the
  terms from each of the 7 workloads described above, but the rules synthesized
  contain more duplicates and overly complicated rules
  (e.g., \C{(- (- b a) (- b a)) \rewritesto\, (sin PI)} instead of \C{0 \rewritesto\, (sin PI)}).
With \enumo, it is easy to design small workloads and compose
  rulesets incrementally.

\subsection{Developing custom workloads for a Halide-inspired domain}
\label{subsec:halide-appendix}
It is possible to \textit{overfit} an \enumo workload by only enumerating terms
  that correspond to the left- and right-hand sides of a desired rule.
For example, creating a workload consisting only of the terms
  \C{(- (* x y) (* z y))} and \C{(* (- x z) y)} in
  order to synthesize the rule \C{(- (* x y) (* z y)) \rewritesboth\, (* (- x z) y)}
  is possible, but requires that the desired rule is known beforehand.
However, the real value of \enumo is its ability to discover useful rewrite rules
  \textit{without} the domain expert needing to already know what rules they
  are looking for.
\enumo allows domain experts to guide the term enumeration, enabling \enumo
  to scale past tools that rely on exhaustive term enumeration.
While it is possible to use guided term enumeration to find specific desired rules,
  as described above, it is more useful to use \enumo workloads to encode
  general insights about the domain of interest.
To see how \enumo enables users to encode \textit{broad}, rather than specific,
  domain knowledge, consider the \enumo program for the Halide-inspired
  grammar (\autoref{subsec:halideeval}).
Here, we do not guide the enumeration towards specific rules we want to synthesize,
  but rather we encode intuition about which parts of the search space are worth
  exploring.

First, we group the operators in the domain into semantic categories:
  boolean operators, arithmetic operators, and predicate operators.
Then, we compute an \enumo workload corresponding to exhaustive enumeration
  using each subset of operators, and synthesize rules using each workload separately.
The intuition guiding this strategy is that useful rules are more likely to be found
  within each category of operators than between categories.
Of course, there are also useful rewrite rules that use operators from multiple categories,
  so we also create a workload that enumerates over all of the operators.
However, we are able to use all of the rules we learned previously to shrink the
  \egraph and make rule inference more tractable.
Finally, we enumerate workloads with larger terms, but apply the \enumo{} \textsf{Canon}
  filter to remove terms whose variables do not appear in canonical order,
  again making the search space smaller and more tractable.
Since we have already learned rules that enable reordering
  and re-associating terms, variable order within a term should not matter as much.

The domain expertise we utilize to develop this \enumo program is
  quite general, not overly specific to the rules we intend to synthesize.
However, it is still valuable to guide the search space in broad ways,
  which is not feasible with one-shot theory explorers like \ruler.
By separating the rule inference problem into smaller searches over subsets of
  the enumeration space, we are able to synthesize a more powerful ruleset
  than prior work.

\section{Discussion, limitations, and future work}
\label{sec:lim}

\Eqsat engines mitigate the effects of
  rule ordering by non-destructively
  applying all the rules in the ruleset
  in each iteration of the algorithm.
However, in practical applications
  where saturation is unlikely,
  the engine completes only a bounded number of iterations
  before resource limits force termination.
Deciding \textit{which}
  rules to run \textit{when}
  becomes critical---splitting
  rules into batches can
  drastically alter the results.
Based on preliminary experiments,
  we find that certain strategies
  can dramatically improve the results of \eqsat tools.
One such strategy is a
  \textit{saturating scheduler}, which
  iteratively
  (1) applies saturating rules
  (\T{is\_saturating} in \autoref{fig:ruleops})
  until they reach saturation, then (2) applies
  the other rules for a single
  iteration; another is the use of
  operators like \T{compress}.

These observations call for a
  more systematic investigation
  of scheduling techniques than this paper provides, but we
  are excited to further explore
  rule scheduling in the context of \eqsat.
We have also only partially explored conditional rule inference in \enumo.
To use \enumo for inferring state-of-the-art
 rules in more complex domains like LLVM IR,
robust conditional rule inference as well as
program analyses to satisfy side conditions will be necessary.

While in this paper we used \enumo to synthesize rewrite
rules for and by equality saturation, the DSL itself is
generic and not restricted to applications in equality saturation.
\enumo lays the groundwork for future applications in
bounded model checking and sketch-guided synthesis.
For example, \enumo could be useful in axiom synthesis tools such as
  LAS~\citep{oopsla22-logic}, where the enumeration order has a large impact on the
  quality of results.

\section{Related work}
\label{sec:related}

Several tools use \egraphs for rule inference~\citep{sat19, ruler}.
\citet{sat19} use enumerative
  synthesis to infer axioms for the CVC4 theorem prover.
Ruler~\citep{ruler} outperformed \citet{sat19}
  in various domains.
  In this paper,
  we show that we can outperform Ruler in terms of both
  scalability and generality of domains (\autoref{sec:eval}).

Similarly, theory exploration is a well-studied topic
 focusing on eagerly synthesizing
 lemmas that may be useful for
 verification tasks.
A recent tool in this space is
  TheSy~\citep{cav21}, which performs 
  inductive theory exploration using
  \eqsat and symbolic values
  to efficiently filter candidate conjectures.
TheSy's key insight is to leverage congruence closure
 to implement an induction prover within the
 \eqsat framework.
Other, similar tools for theory exploration
 use random testing to find
 potential candidates~\citep{hipspec, quickspec}.
IsaCoSy~\citep{isacosy} synthesizes inductive
  theorems for the Isabelle theorem prover~\citep{isabelle}.
To keep the search space of terms tractable,
  IsaCoSy selectively enumerates only terms that
  are not reducible from existing rules.
A similar technique is used by \citet{seplogic}
  in lemma synthesis for proving
  entailments with separation logic.
We believe the abstractions provided by \enumo for
  guided search and ruleset manipulation
  can be used to scale lemma synthesis
  in these tools.
In future work, we would like to express the
inductive prover from  TheSy
 in \enumo.

Many custom tools have been proposed
  for synthesizing rewrite rules in
  specific domains.
\citet{queso} proposed a tool that
  synthesizes rewrite rules for
  quantum circuit optimization.
\citet{taso19} developed a tool synthesizing
  graph substitutions for deep neural networks.
RuleSy~\citep{rulesy} uses a combination of
synthesis and specification mining to find
proof rules representing the mined
specification.
\citet{wetune} presented a tool for
  discovering and verifying query rewrite rules.
In contrast,
  \enumo's DSL-based approach is
  not specialized to any particular domain;
  our evaluation in \autoref{sec:eval}
  shows that \enumo works across diverse domains.
Prior work has used machine learning to assist in
  rewrite rule inference~\citep{oopsla22-logic, swapper};
in particular, \citet{swapper} supports some
  forms of conditional rules.
More recently, large language models have shown promise
  for synthesis tasks~\citep{hysynth, asap, shraddha};
  in \autoref{sec:case-study-llm}, we show that \enumo's
  operators make it easy to incorporate LLMs into
  theory exploration, where they complement guided search.
This paper shows that using \enumo's novel term enumeration
  primitives, rule inference scales
  to support grammars that have conditional
  operators; however,
  full support for conditional rule inference
  is left for future work.

Finally, several tools have focused on automatically inferring
  peephole optimizations~\citep{bansal, alive, fraser, optgen}, and
  instruction selection~\citep{buchwald}.
Two major challenges with these optimizations
  are the presence of side conditions
  and their large grammars.
This paper shows that
  with \enumo's guided enumeration strategy,
  it is possible to find rewrite rules with 
  side conditions.
  We also show that it is possible to
  scale to
  large grammars, like that of Halide~\citep{halide}.
We will continue to explore
  better support for more general conditional
  rewrite rule inference,
  and we are excited to use \enumo
  to infer more optimizations for frameworks like LLVM.

The \enumo DSL is designed to facilitate
  efficient term enumeration given a grammar.
Effective enumeration has been explored in many other contexts like
  relational algebra, sorting algorithms, testing, and generating well-typed
  lambda terms~\citep{alice, icfp16, feat, flajolet, welltyped}.
In the most closely related work, \citet{feat}
  propose \textit{Feat}, a Haskell library for composing enumerations.
They use a lazy mechanism (\textit{functional enumeration})
  to scale enumeration and leverage memoization to efficiently index into a stream
  of enumerated terms.
In a previous prototype of \enumo, we explored a similar mechanism,
  but we found that in the context of rewrite rule inference,
  enumeration time is never the bottleneck---working
  with a large \egraph is.
Therefore, in our final prototype, we use a simpler method for materializing
  a workload into a concrete set of terms.
A similarity between the \enumo DSL and
  the Feat library is the idea of
  composable workloads.
\citet{feat} define a set of combinators that
  allow them to compose smaller
  enumerations effectively.
As \autoref{sec:slide} showed,
  \enumo workloads
  can be composed using
  a set of operators (\Plug, \Filter, \Union), some of which
  are similar to Feat's (e.g., \textit{union}ing two workloads or enumerations).
A key feature of \enumo is that it
  only evaluates a workload when
  converting it to an \egraph (as shown in \autoref{sec:slide})---this
  allows \enumo to leverage a unique set of operators like
  \C{plug} and \C{filter} to
  optimize a workload before it is evaluated
  and converted to an \egraph.

\section{Conclusion}
\label{sec:conclusion}
This paper presents \enumo, a new domain-specific language
  for rewrite rule inference using \eqsat.
\enumo offers novel term enumeration
  primitives and exposes useful ruleset operators
  that enable incremental, composable,
  workload-guided rewrite rule inference.
We also introduce a
  new \textit{\ff} algorithm for generating
  rewrite rules;
  \ff finds rewrite rules for domains not supported by
  prior tools.
\enumo also subsumes the capabilities of
  state-of-the-art tools for
  rule inference~\citep{ruler, sat19} in terms of
  ruleset quality and scalability.
Several case studies demonstrate that small, modular
  \enumo programs
  generate useful rulesets that can be plugged
  into existing \eqsat tools
  or composed to quickly find rulesets across diverse domains.
\enumo is also simple to extend:
  adding two operators sufficed to incorporate
  large language models into theory exploration,
  where they complement \enumo's guided search.
\enumo enables users
  to strategically guide the rule inference process at a high level, and
  to incrementally build effective rulesets.

\begin{acks}
  We thank the anonymous reviewers for their thoughtful feedback.
  We are grateful to Ian Briggs and Pavel Panchekha for helping us run
  Herbie and Megalibm using \enumo's rules.

  This material is based upon work supported by the National Science
  Foundation under Grant Nos.~2232339 and 2312195, and by the NSF
  Graduate Research Fellowship Program under Grant No.~DGE-2140004.
  Any opinions, findings, and conclusions or recommendations expressed
  in this material are those of the author(s) and do not necessarily
  reflect the views of the National Science Foundation.
\end{acks}

\section*{Data availability statement}
  The code and benchmarks for this paper's original evaluation are
  archived as the OOPSLA 2023 artifact on Zenodo
  (\url{https://doi.org/10.5281/zenodo.8140951}).
  \enumo{} is developed openly on GitHub
  (\url{https://github.com/uwplse/ruler}).
  The code and results for the new case study extending \enumo with
  LLM-based operators (\autoref{sec:case-study-llm}), and the code
  behind \autoref{table:herbie}'s baselines, are at tag \C{jfp2026}
  of that repository
  (\url{https://github.com/uwplse/ruler/tree/jfp2026}).

\bibliography{reference}

\end{document}